    \documentclass[reprint,aps,floatfix]{revtex4-1}
                   \usepackage{dcolumn}

\usepackage{amsmath}
\usepackage{graphicx}
\usepackage{bm}
\usepackage{color}

\newcommand{\GeV}{\makebox{ GeV}}
\newcommand{\beq}{\begin{equation}}
\newcommand{\enq}{\end{equation}}
\newcommand{\beqa}{\begin{eqnarray}}
\newcommand{\beqast}{\begin{eqnarray*}}
\newcommand{\enqa}{\end{eqnarray}}
\newcommand{\enqast}{\end{eqnarray*}}

\def\GeV{\nobreak\,\mbox{GeV}}

\begin{document}

\title{ Analytical representation for amplitudes and differential cross section of pp elastic scattering at 13 TeV }
\thanks{Corresponding author. Email: erasmo@if.ufrj.br\vspace{6pt}}  
\author{E. Ferreira$^{\rm a}$}
\author{A. K. Kohara$^{\rm b}$}
\author{T. Kodama$^{\rm a,\, c}$}

\begin{abstract}

With  analytical representation for the  pp  scattering amplitudes  introduced and tested at        
lower energies, a description  of high precision is given of the $d\sigma/dt$  data at 
$\sqrt{s}$= 13 TeV for all values of the momentum transfer, with explicit identification  
of the real and imaginary parts. In both $t$ and $b$ coordinates the amplitudes have terms 
identified as of non-perturbative and perturbative nature, with distinction of their 
influences in forward and large $|t|$ ranges and in central and peripheral regions respectively. 
In the forward range, the role of the Coulomb-nuclear interference phase is investigated. 
The energy dependence of the parameters of the amplitudes are reviewed and updated, 
 revealing a possible emergence of a peculiar behavior of elastic and inelastic profiles in 
b-space for central collisions, which seems to be  enhanced quickly at higher energies.  
Some other models are also briefly discussed in comparison, including the above mentioned 
behavior in b-space. 
\keywords{elastic differential cross-section \and  total cross section \and scattering amplitudes}
 \end{abstract}

\affiliation{  
$ ^{\rm a}$ {\em Instituto de F\'{\i}sica, Universidade Federal do Rio de Janeiro   \\
C.P. 68528, Rio de Janeiro 21945-970, RJ, Brazil } \\
$ ^{\rm b}$ {Departamento de Engenharia Qu\'imica, Centro de Tecnologia da Ind\'ustria Qu\'imica e T\^extil, SENAI, Rio de Janeiro 20961-020, RJ, Brazil}  \\
$ ^{\rm c}$ {Instituto de F\'isica, Universidade Federal Fluminense, Niter\'oi 24210-346, RJ, Brazil}  \\}

\keywords{pp elastic scattering;  hadronic amplitudes }




\maketitle



\section{Introduction\label{intro}}

\begin{figure}[b]
  \includegraphics[width=8cm]{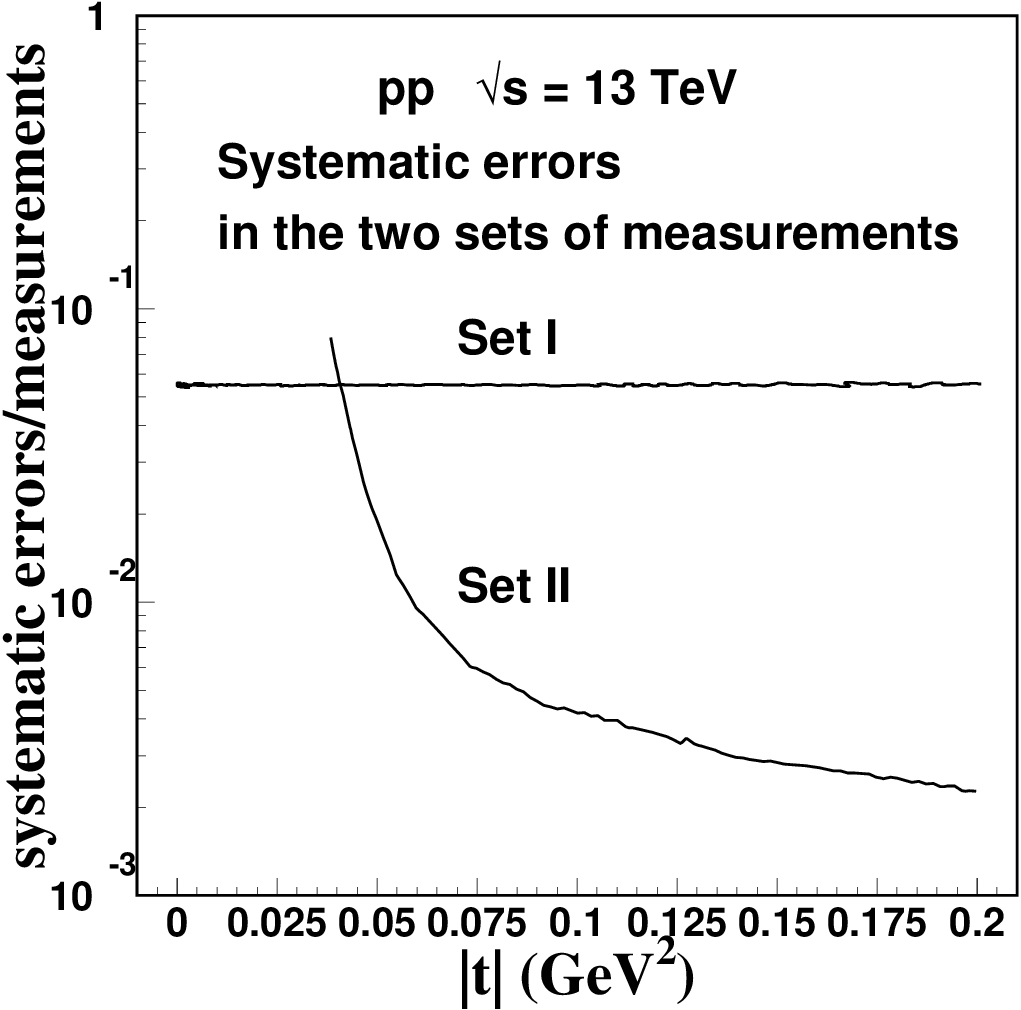} 
\caption{Systematic errors in data  Set I \cite{Totem_13_1} and Set II \cite{Totem_13_2}. In the  
$|t|$ range with superposition  ($0.038400 \leq |t| \leq 0.201041$) it seems that in general the 
data in Set II may  be considered as more reliable (1\% systematic errors), except 
 for the first 11 points. Set I has 5\% systematic error bars.}
\label{errors-fig}
\end{figure}

Totem Collaboration in LHC has produced two sets of data data on elastic pp scattering at $\sqrt{s}$=13 TeV in  
separate publications  \cite{Totem_13_1,Totem_13_2,Totem_13_3}, covering the following $|t|$  ranges 
\begin{itemize}
  \item Set I -  $|t|= [0.000879 - 0.201041]~ \GeV^2 $ , with N=138 points \cite{Totem_13_1} ;
\item  Set II -  $|t|= [0.0384 - 3.82873] ~  \GeV^2 $ , with N=290 points \cite{Totem_13_2} .
\end{itemize}  
With respect to systematic errors, the two sets of measurement are presented with very different 
features: errors of about 5$\%$ for I  and less than  1$\%$  (except for the first 11  points) for 
Set II. The situation, illustrated in Fig.\ref{errors-fig}, influences the analysis of the data. 
The very large systematic errors in Set I indicates the necessity of special care on  its 
use for the determination of the forward scattering structure. 

There are 56 points of small $|t|$  in Set I, up to $|t|=0.037335\,\GeV^2$, where Set II starts,
and thus there is a basis of 56 + 290 = 346 data  points to perform  a global description of the 
13 TeV data. We also build   a combined file merging the points of the common range, with a total 
of 138+290=428 points that are used in an overall test.

The data  of  Set I have been  studied  \cite{PLB2019} with  forms of amplitudes restricted 
to small $|t|$ values. The treatment of this range requires detailed  account of  the 
Coulomb-nuclear interference, and it was shown  that  the model-independent determination  
of the amplitude   in these representations is unreliable with the present data alone, due to 
the small value of the $\rho$ parameter  and to the  assumption of a model for the   
treatment of the Coulomb-nuclear interference phase that needs to be tested at such high energies. 
In the forward direction the real part contributes to only about 1\% of the observed $d\sigma/dt$, 
and it is necessary to have a well-inspired extraction of the imaginary part, requiring  data 
of very regular behaviour, to allow the determination of the properties of the real part, 
 such as the $\rho$ parameter and the amplitude slope.

\begin{figure}[b]
  \includegraphics[width=8cm]{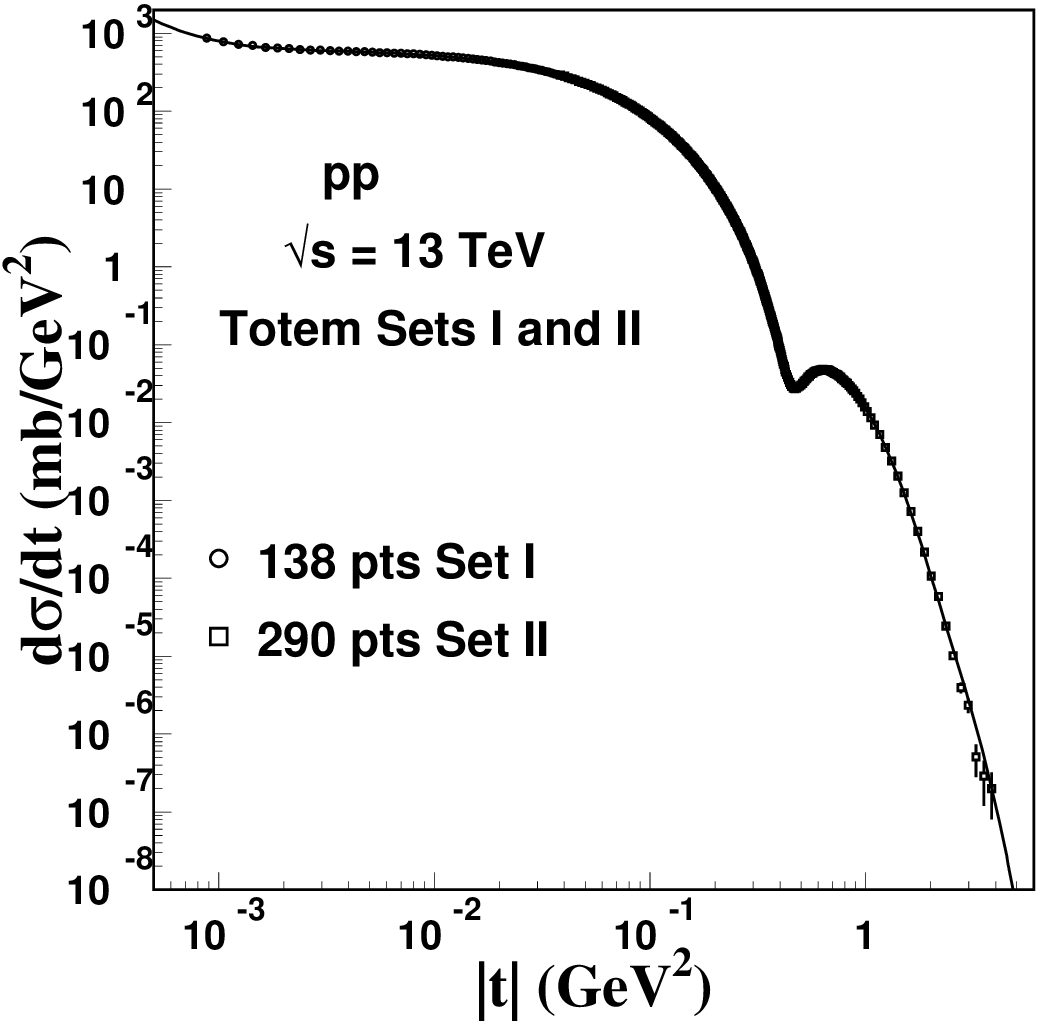} 
\caption{Analytical representation of all  data  points of Totem measurements  at 13 TeV
  \cite{Totem_13_1,Totem_13_2}, using 4 adjusted 
parameters  \cite {KFK_1,KFK_2,KFK_3,flavio1} for each of the real and imaginary parts.
The total of 428 data points is described with $\chi^2=1.567$ (statistical and systematic 
errors added in quadrature) and $\chi^2=5.186 $ (statistical errors only).
Details are given in  Secs.\ref{KFK_section} and \ref{data_analysis}. }
\label{all_TOTEM_data}
\end{figure}

 Putting all information together, we achieve  a unified  treatment of 428 data points, 
identifying analytically the real and imaginary parts (with 4 parameters each)  of the complex 
elastic amplitude, with   remarkable values $\chi^2 = 1.567$ with statistical and systematic errors 
added in quadrature and  $\chi^2 = 5.186 $ calculated with statistical errors only. 
Everywhere in  the present text $\chi^2$  is a short for $\chi^2/d.o.f.$.  
The graphical representation of this result is shown in Fig.\ref{all_TOTEM_data}.     
  The present treatment is similar to previous work that was   very effective 
at lower energies  1.8 - 1.96 TeV of Fermilab \cite {KFK_1} and 7-8 TeV of 
LHC \cite{KFK_2,KFK_3}.

The   large  $|t|$ range of Set II is coupled   sensibly with the (energy independent) tail
of perturbative three-gluon exchange observed at $\sqrt{s}=27.4 ~ \GeV $ \cite{Faissler}, 
with 39 points in  the range  $ 5.5 \leq |t| \leq 14.2 ~ \GeV^2$. 
The first identification of the  energy independence of the $d\sigma/dt$  behaviour for 
large $|t|$ in pp elastic scattering 
was made in the comparison of data at $\sqrt{s}$ = 19.6 and 27.4 GeV \cite{Conetti}. 
The theoretical explanation for the $1/|t|^8$ behaviour of $d\sigma/dt$ for large $|t|$ 
  in terms of the real three-gluon exchange  amplitude  was given by Donnachie 
and Landshoff \cite{DL1}. 
The universality  is demonstrated for   energies below 
$\sqrt{s}$ = 62.5 GeV in   Fermilab and  CERN/ISR  measurements \cite{flavio1,flavio2}
  (see figures  in these two papers), 
showing  smooth connection between the   range 
of small and mid-$|t|$ combining perturbative and nonperturbative terms  and the range 
of large $|t|$ of FNAL \cite{Faissler} measurements  dominated by three-gluon exchange. 
The  role of the real    amplitude in 
the large $|t|$ sector of pp elastic scattering is then confirmed. 
 
 The transition range  from   2 to 5 $\GeV^2$  gives information on the magnitude 
and sign of the real part of the hadronic amplitude, that is dominant for large $|t|$. 
 Unfortunately 
the  LHC pp measurements  at 7 and 8 TeV \cite{KFK_2,KFK_3} are restricted to $|t|$
less than  $ 2 ~ \GeV^2$, and the connection between mid and large $|t|$ regions 
remained in the non-quantitative level, although there is   clear indication , as 
shown in   Fig.6 of the 7 TeV paper  \cite{KFK_2},
where   the data at 52.8 GeV and 7 TeV  are exhibited.
 At 13 TeV the  measurements reach  almost $|t|= 4 \GeV^2$,  
allowing    investigation in an important  extended range.  
Using the same representation described above,  
with a proper connection between  the 13 TeV and the 
17.4 GeV data, we obtain an analytical  form  embracing 467 (= 428+39) 
 data points, with $\chi^2 = 1.731$ and $\chi^2 = 5.042$ using total errors 
(combined statistical and systematic) and pure statistical errors respectively.

The present work  uses the amplitudes introduced in previous papers \cite{KFK_1,KFK_2,KFK_3}, 
  expressed  in both $t$ and $b$ coordinates, with explicit forms  
    for the real and imaginary amplitudes: the disentanglement 
of the two parts  is essential for the description of the dynamics of the process.
 The superposition of non-perturbative and perturbative terms 
in both real and imaginary parts produces remarkable structure in the elastic 
differential cross section that   faithfully reproduces the data. 
In the following this framework is referred to as KFK model.  

 In   Sec.\ref{KFK_section}   we review the construction of the amplitudes in the KFK model,
   inspired on the early applications of the  Stochastic Vacuum Model (SVM) to high-energy 
elastic scattering. The $b$ and $t$ space coordinates are analytically related, with terms 
representing perturbative and non-perturbative dynamics. 

In Sec.\ref{data_analysis}  
we apply the KFK amplitudes to describe in detail  the forward, mid and large $|t|$  ranges,
obtaining a unique solution valid with high precision for all $|t|$, as shown in  
Fig.\ref{all_TOTEM_data}. Separate  attention is given to an extension of the representation 
to the range of high $|t|$ measured at 27.4 GeV \cite{Faissler} and also to the small $|t|$ 
range of Set I re-examining the role of the Coulomb-nuclear interference phase \cite{PLB2019}.
 In Sec.\ref{amplitudes}, the properties of the amplitudes in $|t|$- and $b$-  coordinates are 
described and discussed  in  separate subsections.
In Sec.\ref{energy} we insert the results of the present analysis at 13 TeV  in the previous 
study of energy dependence of the  KFK  framework, updating  description and 
predictions. There, we report a new  behaviour of the profile functions in b-space in the domain 
of central collisions, which seems to be enhanced quickly at high energies. This 
observation was not possible without the present 13 TeV data.  Sec.\ref{MODELS} compares our 
description 
with other models  and  Sec.\ref{CONCLUSIONS} presents remarks and critical evaluation.
 
\section{KFK model : Analytical Representation of the amplitudes \label{KFK_section} }

 The Stochastic Vacuum Model(SVM) is based on the functional integral approach 
\cite{Nachtmann1} to high energy scattering  that relates high energy scattering with 
nontrivial properties of QCD vacuum \cite{Dosch,Dosch_Simonov}. 
 The central element is the gauge invariant Wegner-Wilson loop, and physical quantities are 
obtained from the vacuum expectation values of the correlations of two loops, 
defined in terms of coordinates  in the transverse collision plane.  
Assuming dominance of Gaussian fluctuations in the field strengths, the calculation   
becomes fully analytical. 
  Observables are written in terms of physical quantities: the value of the gluon 
condensate,that determines the strength of this non-perturbative dynamics, 
 and the correlation length, that is the parameter  of the  
loop-loop correlation function that sets the scale  for the geometric 
dependence  in $b$-space. These quantities have values fixed by 
hadronic properties and by lattice calculations  \cite{DiGiacomo,Meggiolaro}. 
With analytic continuation from Euclidean to Minkowski  space \cite{SVM3}
 gauge-invariant  dipole-dipole scattering  is constructed.  

 The amplitude of  non-perturbative  hadron-hadron  scattering 
in the eikonal  approximation is factorized  with the product of the correlation 
of  loops (representing elastic scattering of two colour dipoles) and the 
factor with the dipole contents in the light-cone wave functions of the 
colliding hadrons \cite{SVM,SVM1,SVM2}. The overlap of the loop-loop
correlation with the hadronic wave-functions of finite size  leads 
to structure of profile function where the basic   correlation
parameter becomes spread, appearing  with effective value that  depends 
on the hadronic sizes  and, in case of scattering amplitudes, can also 
be modified by the collision energy.  These effective representations 
of the correlations proper of the QCD vacuum are not expected 
to be very different from the static lattice  determination. 

 Besides hadron-hadron scattering, the concept of the 
loop-loop correlation was also applied   to the non-perturbative 
exclusive photo- and electroproduction of vector mesons \cite{SVM2,Photo,Electro}. 
  
 The KFK  model writes   analytical  forms for the pp and ${\rm p \bar p}$  
elastic scattering amplitudes in $t$ and  $b$ spaces,    
 based on previous experience with the Stochastic Vacuum Model (SVM)    
\cite{SVM}, using a scale (correlation) length parameter  and the asymptotic 
(large $b$) behaviour  of the profile function as guiding ingredients.  

KFK model introduced non-perturbative and perturbative contributions  
\cite{flavio1,flavio2}, later assumed as necessary long and short range 
terms in the loop-loop correlation \cite{SVM2}. 
 The effective gluon mass  introduced to 
control the infrared range in the perturbative  correlator  enters 
in the overlap product with the proton dipole content  and appears 
in the profile function in KFK through a  simple 
Gaussian  term  as in Eq.(\ref{b-AmplitudeN}). 
 
The T-matrix element in SVM is purely imaginary, and with missing real part     
 $d\sigma/dt$  cannot  be calculated in the full $|t|$ range.  
  KFK introduces a real part that is a mirror image of the imaginary 
amplitude.   The real part  
is dominant for large $|t|$, and has crucial role  in the dip-bump  region 
of pp elastic scattering around  0.4-0.5 $\GeV^2$ where the imaginary part passes 
through zero. The sophisticated  dip-bump  structure in  $d\sigma/dt$ 
 requires   delicate property of  the real part valid   in this range.   Both 
parts must have perturbative and non-perturbative terms, and must have zeros,
signs and magnitudes following  theoretical principles and reproducing observations
\cite{flavio1,flavio2}. 
The  zero in the real part at small $|t|$  predicted by a theorem by 
A. Martin \cite{Martin}, is confirmed with the LHC data \cite{MURILO}. 
while  the imaginary part has a zero   responsible for the dip-bump structure 
 in $d\sigma/dt$.  
  
 The analytical forms proposed for the non-perturbative terms of the amplitudes  
are inspired in the behaviour of the profile function for large  $b$ found in the 
calculation with SVM \cite{flavio1,SVM}, with a combined exponential-Yukawa 
dependence. The  Fourier transforms to $t$-space  present features that  can 
effectively represent the data for all $|t|$. 
 As $b$ is not an observable quantity, the  construction 
is tested in $|t|$  space, and   parameters  are  fixed   by  experiments.
Accurate description of the data is obtained with four  parameters in each part 
of the complex amplitude.    

The disentanglement of the two parts of the complex amplitude is not at all  
trivial. The connection  with the three-gluon exchange contribution  helps in the  
identification of the sign  and magnitude of the real part, and an additional 
term for perturbative three-gluon exchange  is introduced separately.  

 The   KFK  model has been investigated  at several energies, 
and   the energy dependence of the parameters comes out  smooth, with
simple parametrization \cite{KFK_1,KFK_2,KFK_3}.  

\subsection{ Impact parameter representation      \label{impact_par}     }

The amplitudes in the Stochastic Vacuum Model \cite{SVM} are originally 
constructed through  $b$-space profile functions, that give insight for 
geometric aspects of the collision, playing role in the eikonal representation, 
where unitarity constraints have  interesting formulation. 
  The  dimensionless $(s,b)$ amplitudes of the pure  nuclear interaction  are 
written in the form 
\begin{equation}
\label{b-AmplitudeN}
\widetilde{T}_{K}(s,\vec{b})=\frac{\alpha _{K}}{2\beta _{K}}e^{-{b^{2}}/{%
4\beta _{K}}}+\lambda _{K}\widetilde{\psi }_{K}(s,b)~,  
\end{equation}%
with a Gaussian term meant to be of perturbative nature and 
a characteristic non-perturbative shape function 
 \begin{equation}
\label{Shape-b}
\widetilde{\psi }_{K}(s,b)=\frac {2e^{\gamma _{K}-\sqrt{\gamma _{K}^{2}+{b^{2}}/{a^2}}}}
{a^2\sqrt{\gamma _{K}^{2}+{b^{2}}/{a^2}}}
\Big[1-e^{\gamma_{K}-\sqrt{\gamma _{K}^{2}+{b^{2}}/{a^2}}}\Big]~.  
\end{equation}
The label $K$ $=R,I$ indicates either the real or the imaginary part of the
complex amplitude. 

The quantity $a$, called correlation length, represents properties 
of the QCD vacuum, where it sets the scale for the loop-loop correlation,
with determination in static (Euclidean space) lattice calculation \cite{DiGiacomo} as 
0.25-0.30 fm. 
After analytic continuation to Minkowski space and overlap with the hadronic 
wave functions,  the non-perturbative 
scale appears  in profile functions of hadron-hadron scattering,
with effective value  modified inside this range. 
In the present work for pp scatering at 13 TeV we find the    value
\begin{equation}
    a^2  =   2.1468 \pm 0.0001 ~ \GeV^{-2}   
     =  (0.2891 \pm 0.0002 ~ {\rm fm} )^2 .   
\end{equation}
The parameters $\alpha_K(s), \beta_K(s), \lambda_K(s)$ with units in $\GeV^{-2}$ 
and $\gamma_K(s)$ dimensionless are functions of the energy. They are determined 
for $\sqrt{s}=13$ TeV with high precision  in Sec.\ref{data_analysis}, 
leading to explicit analytical  expressions for the imaginary and real amplitudes.
The Gaussian form of the first term in Eq.(\ref{b-AmplitudeN}) corresponds to 
the perturbative part of the loop-loop correlation introduced in developments 
of  SVM, following results suggested by lattice calculations.  The second term, 
referred to as shape function, corresponds to  contributions from non-perturbative  
loop-loop correlation function. It is zero at $b=0$, $\widetilde{\psi }_{K}(s,b=0)=0$,   
and  is  normalized as 
\begin{equation}
\frac{1}{2\pi }\int d^{2}\vec{b}\ \tilde{\psi}_{K}\left( b,s\right) =1~.
\label{psinorm}
\end{equation}%

   Eq.(1) represents a parametrized  formulation of the profile function 
based on the  SVM proposal. The perturbative and non-perturbative terms of the 
amplitudes are dominant for small and large $b$ respectively. For large $b$,  
corresponding to peripheral collisions, the amplitudes fall down with a
exponential-Yukawa-like tail, 
\begin{equation}
\sim \frac{1}{b}e^{-b/b_{0}},
\end{equation}%
that reflects the correlations of  loops  at large distances. This 
asymptotic behaviour inspired the construction of the shape function  
 $ \widetilde{\psi }_{K}(s,b)$   for  Eq.(\ref{b-AmplitudeN}). 
 
\subsection{ $t$-space representation       \label{tspace_par}   }

In the classical limit the variable  $b$ is connected with the 
impact parameter, but it is not directly observable, and the treatment 
of data is made in $(s,t)$ space.  One advantage of the shape function in KFK 
is that there is  explicit analytic Fourier 
transformation for the amplitudes in Eqs.(\ref{b-AmplitudeN},\ref{Shape-b}), 
so that the scattering properties can be studied directly in both frameworks.  
 
In our normalization the elastic differential cross section is written 
\begin{eqnarray}  \label{Sigma_diff}
\frac{d\sigma(s,t)}{dt}&=& (\hbar c)^2[T_I^2(s,t)+ T_R^2(s,t)] \\
&=& \frac{d\sigma^I(s,t)}{dt} + \frac{d\sigma^R(s,t)}{dt} ~ ,  \notag
\end{eqnarray}
with  $T_{R}(s,t)$ and $T_{I}(s,t)$ in $\GeV^{-2}$ units, and 
$$  (\hbar c)^2~ =  ~ 0.389379 ~{\rm{mb}}\GeV^2 ~ . $$
   The complete  amplitudes ,
contain the nuclear and the Coulomb parts as 
\begin{equation}
T_{R}(s,t)=T_{R}^{N}(s,t)+\sqrt{\pi }F^{C}(t)\cos (\alpha \Phi )~,
\label{real}
\end{equation}%
and 
\begin{equation}
T_{I}(s,t)=T_{I}^{N}(s,t)+\sqrt{\pi }F^{C}(t)\sin (\alpha \Phi )~,
\label{imag}
\end{equation}%
where $\alpha ~$is the fine-structure constant, $\Phi (s,t)$ is the 
interference  phase (CNI)  and $F^{C}(t)$ is related with the proton form factor 
\begin{equation}
F^{C}(t)~=(-/+)~\frac{2\alpha }{|t|}~F_{\mathrm{proton}}^{2}(t)~,
\label{coulomb}
\end{equation}%
for the pp$/$p$\mathrm{{\bar{p}}}$ collisions. The proton form factor is
taken as%
\begin{equation}
F_{\mathrm{proton}}(t)=[t_{0}/(t_{0}+|t|)]^{2}~,  \label{ff_proton}
\end{equation}%
where $t_{0}=0.71\ $GeV$^{2}$.  

We  recall the new measurements of the  proton radius \cite{proton_radius}
and changes in the proton form factor \cite{proton_formfactor}.
These changes in the electromagnetic and hadronic structure of the proton 
may become  important for the analysis of forward elastic scattering, when their 
quality improves. As it has been proved \cite{PLB2019}, this is not the case at 
the present, and we use the quantities as written above.  

  The expressions $T_R^N(s,t)$ and $T_I^N(s,t)$
represent  the nuclear amplitudes for the   terms   written
in  Eq.(\ref{b-AmplitudeN}).
The non-perturbative shape functions in $t$-space  obtained by Fourier transforms 
 are written 
\begin{eqnarray}
 \label{psi_st}
&&\psi _{K}(\gamma _{K}(s),t) \\
&=&2~\mathrm{e}^{\gamma _{K}}~\bigg[{\frac{\mathrm{e}^{-\gamma _{K}\sqrt{%
1+a^2|t|}}}{\sqrt{1+a^2 |t|}}}-\mathrm{e}^{\gamma _{K}}~{\frac{e^{-\gamma
_{K}\sqrt{4+a^2|t|}}}{\sqrt{4+a^2 |t|}}}\bigg]~,   \nonumber
\end{eqnarray}%
%
with the property 
\begin{equation}
\psi _{K}(\gamma _{K}(s),t=0)=1~   .  \label{psinorm2}
\end{equation}%
 Use is made of   the integration formula 
\begin{equation}
\int_0^\infty J_0(\beta u ) \frac{e^{-\rho \sqrt{\gamma^2+u^2}}}
 {\sqrt{\gamma^2+u^2}} ~ u ~ du = \frac{e^{-\gamma \sqrt{\rho^2+\beta^2}}}
     {\sqrt{\rho^2+\beta^2}} ~ .
  \label{E13}
\end{equation}
 In addition to the Fourier transform of the perturbative part in  Eq.(\ref{b-AmplitudeN})
we introduce in the real part  a term $R_{ggg}\left( t\right) $ 
representing the perturbative three-gluon exchange   \cite{DL1,flavio1} that
  appears in the large $|t|$ region, and the complete  nuclear amplitudes
are then written 
\begin{eqnarray}
\label{hadronic_complete}
&& T_{K}^{N}(s,t)\rightarrow T_{K}^{N}(s,t) \\
&& =\alpha _{K}(s)\mathrm{e}^{-\beta_{K}(s)|t|}+ \lambda _{K}(s)\psi
_{K}(\gamma _{K}(s),t)  \nonumber \\
&& +\delta_{K,R}R_{ggg}\left( t\right) ,~~~K=R,I~,  \nonumber 
\end{eqnarray}
with $K=R,I$, and 
where the Kronecker delta symbol $\delta _{K,R}$ is introduced so that 
$R_{ggg}\left( t\right) $ contributes only to the real part.
 Eqs.(\ref{psi_st},\ref{hadronic_complete}) constitute  the KFK model 
 for the pp and $ {\rm p \bar p} $  elastic amplitudes in $t$ space.

The limits of the amplitudes for small $|t|$ give the total cross section $%
\sigma$(optical theorem), the ratio $\rho$ of the real to imaginary amplitudes 
and the slopes $B_{R,I}$ at $t=0$ through
\begin{eqnarray}
\sigma(s)  
 &=& (\hbar c)^2~ 4\sqrt{\pi} ~ T_{I}^{N}(s,t=0)~ \\
 &=&  4\sqrt{\pi} \left(\hbar c\right)^{2}~[\alpha_{I}(s)+\lambda_{I}(s)]~  \nonumber \\
&=& ~ 2.7606 ~[\alpha_{I}(s)+\lambda_{I}(s)]~ {\rm mb }~,  \nonumber
\label{sigma_par}
\end{eqnarray}%
\begin{equation}
\rho(s)=\frac{T_{R}^{N}(s,t=0)}{T_{I}^{N}(s,t=0)}=\frac{\alpha_{R}(s)+%
\lambda_{R}(s)}{\alpha_{I}(s)+\lambda_{I}(s)}~  \label{rho_par}
\end{equation}
and
\begin{eqnarray}
 B_{K}(s)&& =\frac{2}{T_{K}^{N}(s,t)}\frac{dT_{K}^{N}(s,t)}{dt}\Big|_{t=0}%
=~\frac{2}{\alpha_{K}(s)+\lambda_{K}(s)}\times \nonumber\\
&& \Big[\alpha_{K}(s)\beta_{K}(s)+\frac{1}{8}\lambda_{K}(s)a^2 \Big(6\gamma
_{K}(s)+7\Big)\Big]~ .
\label{slopes_par} 
\end{eqnarray}
 
 The   tail term $R_{ggg}\left( t\right) $, producing a universal 
(not energy dependent) $|t|^{-8}$ form for large $|t|$ in $d\sigma/dt$ was studied 
  in the analysis
of the experiments at CERN-ISR, CERN-SPS \cite{flavio1}, 1.8 TeV \cite{KFK_3} and 7 TeV \cite{KFK_1}.
To restrict this contribution to the large $|t|$ region,  we create a connection factor, writing  
 \begin{equation}
R_{ggg}(t)\equiv\pm \frac{d_1}{t^{4}} [1-e^{-d_2(t^2-d_0)}][1-e^{-{\rm x}|t|}]^{d_3} ~, 
\label{R_tail} 
\end{equation}
where the last two factors cut-off  this term smoothly in the domain  from 2 to  5.5 $\GeV^2$, 
and the signs $\pm$ refer to the pp and p$\mathrm{{\bar{p}}}$
amplitudes respectively.
The  detailed form of the factor in Eq.(\ref{R_tail})
   must be adequate for the description of 
the data for  $|t|$ values   in the transition range
connecting  the experimental points \cite{Faissler} at 
$\sqrt{s}= 27.4 \GeV$. 
In  Sec.\ref{data_analysis},  the proposed  parameters are 
\begin{eqnarray}
&& d_0=9~ \GeV^4 ,~ d_1=0.563 \pm 0.008  \GeV^6 ~ , \\
&& d_2
=0.16\pm 0.01  \GeV^{-4} ~ , ~  d_3=48   ~ , ~  {\rm x} =1 \GeV^{-2} ~. \nonumber 
\label{tail_par}
\end{eqnarray}
The peculiar form of Eq.(\ref{R_tail}) is  explained in  Subsection \ref{Faissler}.

\section{  Description  of the 13 ${\bf  \rm  TeV} $ data   \label{data_analysis}  }
 
\begin{figure*}[b]
 \includegraphics[width=8cm]{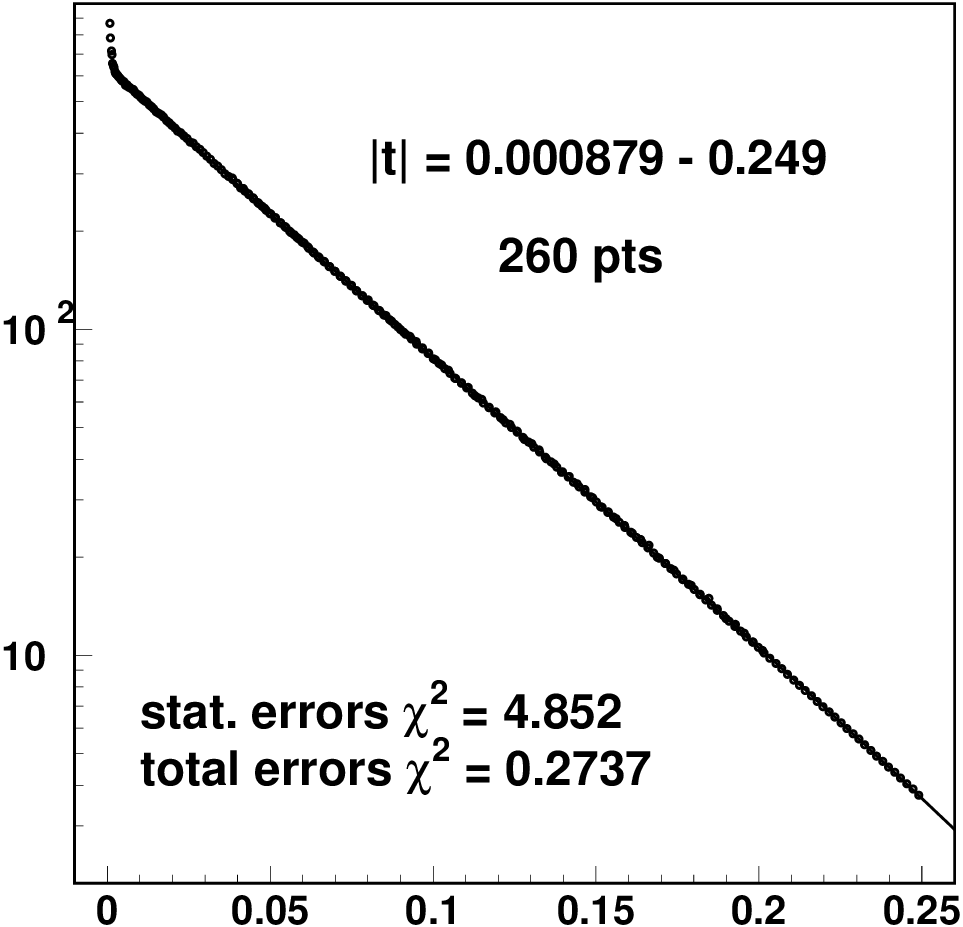} 
 \includegraphics[width=8cm]{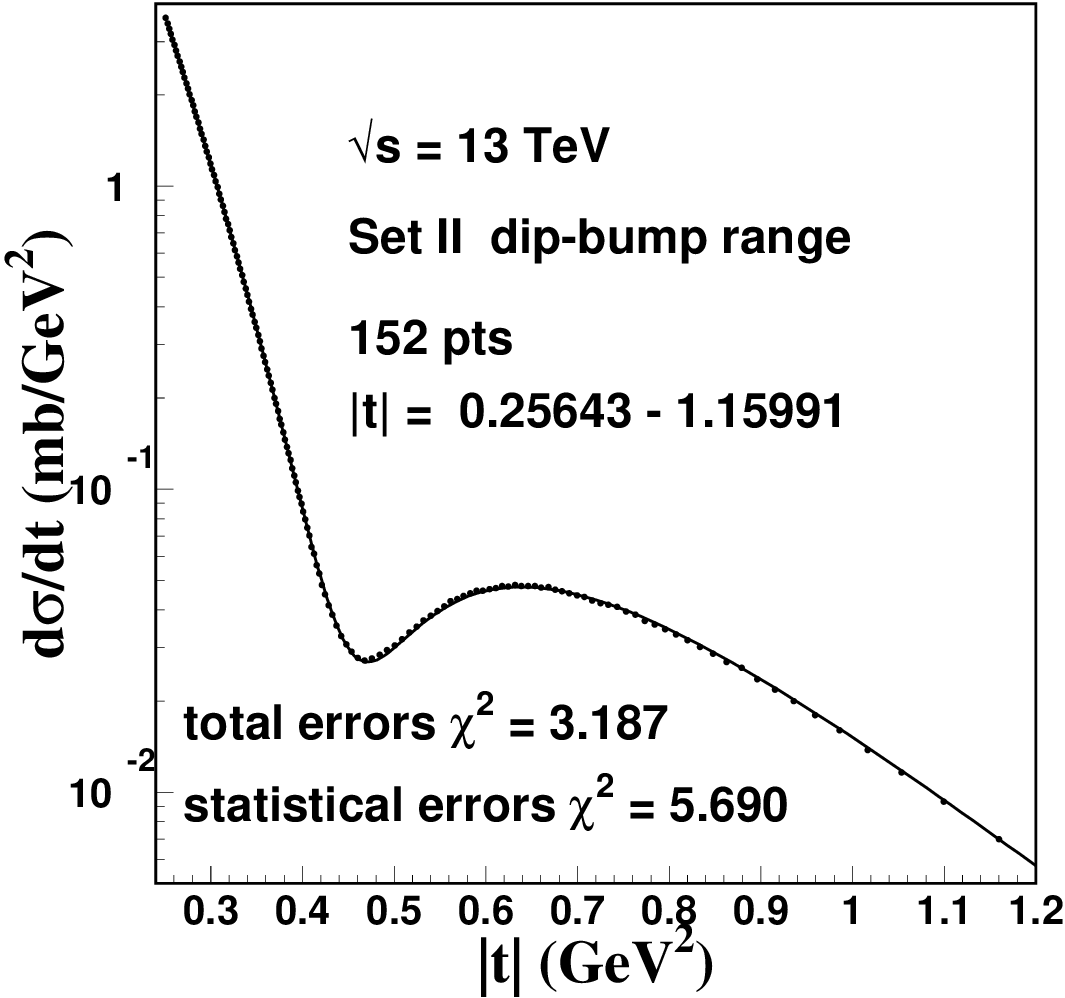}
  \includegraphics[width=8cm]{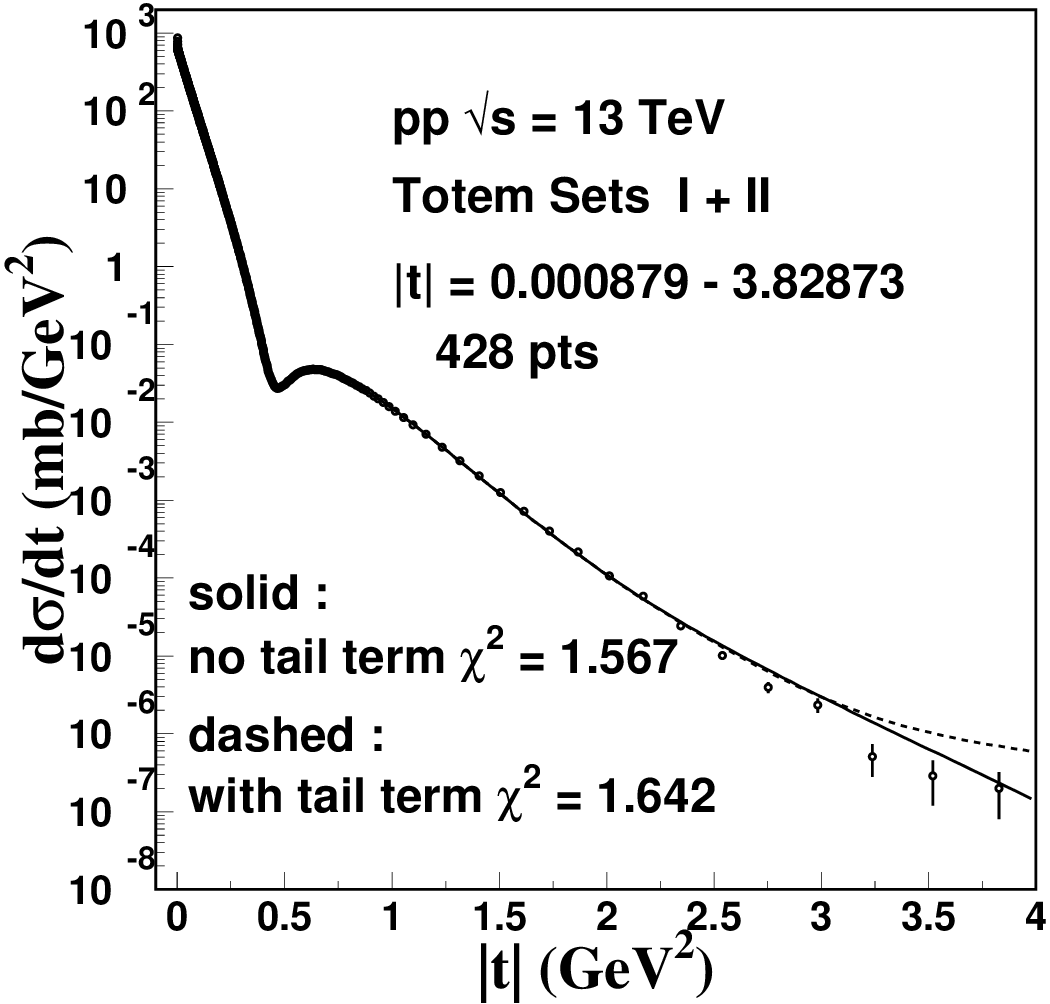} 
  \includegraphics[width=8cm]{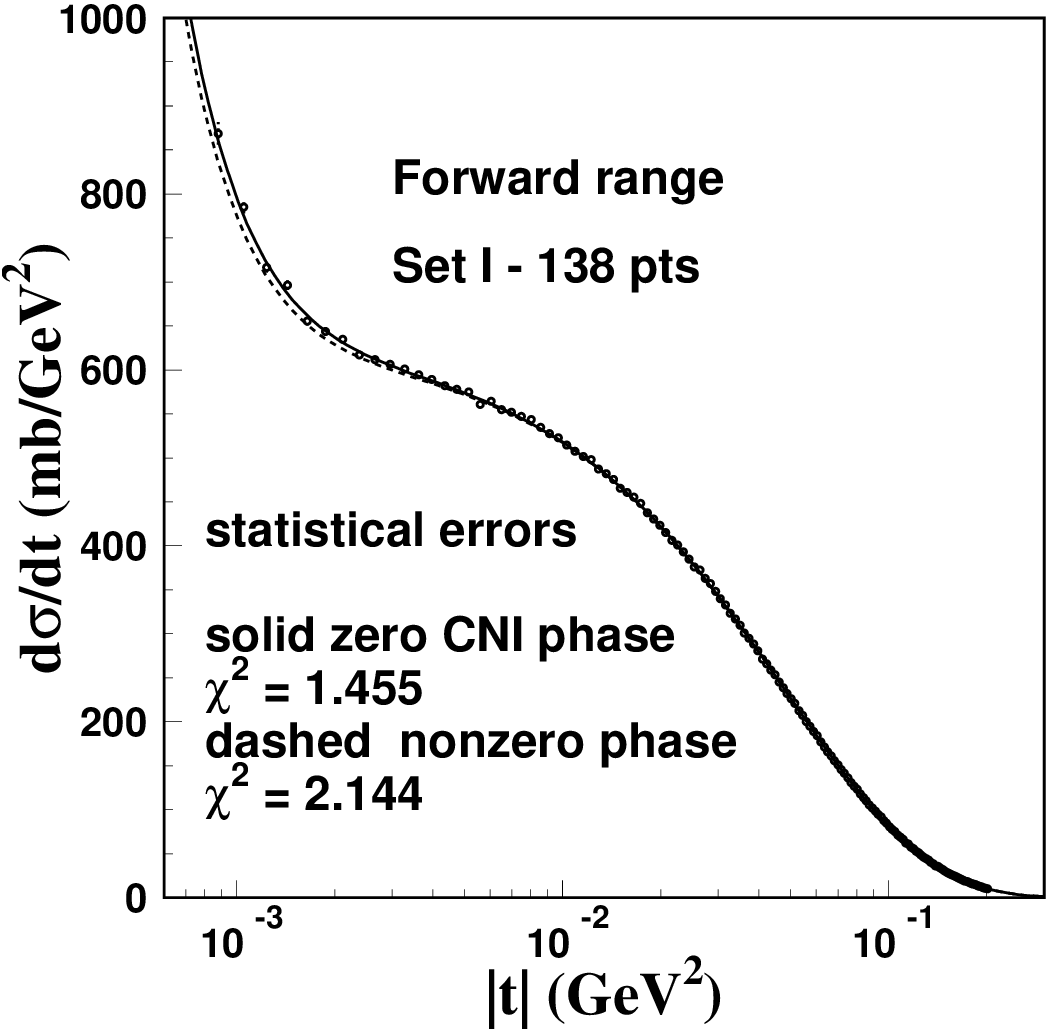} 
\caption{\label{ranges} Representation in the KFK model of  separate $|t|$ ranges of Sets I and  II of 
Totem measurements at 13 TeV, 
with unique analytical form and parameter values given in Table \ref{tableone}. 
Plot c) shows in dashed line the displacement due  to the inclusion of the $R_{ggg}(t)$ term in the amplitude. 
In Fig.\ref{NOVAS} we show how this term implies the connection with the data of large $|t|$ 
at $\sqrt{s}=27.4 \GeV$.  
 In the small- and mid-$|t|$ ranges of plots a) and b) the influence of the tail term is not relevant
 in the plots. 
In  plot d) for small $|t|$ we show lines for calculations with Coulomb-nuclear 
interference phase $\phi$ included in the usual form (dashed line), and   with   phase put as zero (solid line);
 numbers are given in Table \ref{tablefour}. }
\end{figure*}

In this section   we obtain the representation of the data  of Totem experiment 
at 13 TeV through the $t$-space  amplitudes of the KFK model written in 
Eqs.(\ref{psi_st},\ref{hadronic_complete},\ref{R_tail}). Plots in 
 Fig.\ref{ranges}  show separately   forward, mid and full $|t|$  ranges of the  
data of Sets I and  II, described  by a unique solution, with the parameters given 
in Table \ref{tableone}. Table \ref{tabletwo} gives  statistical quantities  
 for different ranges of the data, obtained with the same unique solution. 
Values of $\chi^2$  are given for calculations with statistical errors and 
for total errors combining statistical and systematic errors in quadrature.  
We also inform the  $\chi^2$ value for a combined set of the first 56 points of Set I 
with the 290 points of Set II (total 346 points), avoiding the superposition of 
ranges.  In the last line  of Table \ref{tabletwo} we inform the $\chi^2$ result for 
a set of 467 points joining the 27.4 GeV data  \cite{Faissler}, using the real amplitude 
that includes the $R_{ggg}$ term of 3-gluon exchange as in Eqs.(\ref{hadronic_complete},  
\ref{R_tail}), while  keeping fixed the parameters of Table \ref{tableone}. The  
 connection of the data of these different energies is illustrated  in Subsec.\ref{Faissler}.  
In Subsec.\ref{forward} we present  specific results of an analysis for the forward 
data of Set I. 

Observable quantities and positions of the zeros are given  in Table \ref{tablethree}.
 
{\small
\begin{table*}[ptb]
\caption{Parameters of the amplitudes in the KFK model determined with the 428 points of  Totem 
measurements at 13 TeV. 
  The QCD quantity related to  correlation function is 
 $a^2=2.1468\pm 0.0001~ \GeV^{-2} =  (1.4652 ~ \GeV^{-1}\pm 0.0002)^2=(0.2891 \pm 0.0002 ~ {\rm {fm}})^2 $ ,
where $a$ is called correlation length. 
 The quantities  $\gamma_I$ and $\gamma_R$ characteristic of the non-perturbative shape functions in 
Eq.(\ref{psi_st})  are dimensionless, while $\alpha_K$, $\beta_K$ and $\lambda_K$ have units $\GeV^{-2}$.  
The index  $K$ means $I,R$.  To have all quantities with same dimensions  $\GeV^{-2}$,  we can use
$\eta_K=\gamma_K a^2 $ instead of $\gamma_K$, as in Sec.\ref{energy}.  
 }
   \label{tableone}
   \begin{tabular*}{\textwidth}{@{\extracolsep{\fill}}cccc|cccc@{}}
     \hline
 \hline
   \multicolumn{4}{c} {Imaginary Amplitude} & \multicolumn{4} {c} {~ Real ~  Amplitude ~   } \\ \cline{1-8}
 $\alpha_I$  &$\beta_I$    &$\lambda_I$   & $\gamma_I$   & $\alpha_R$    &$\beta_R$   &$\lambda_R$   &$\gamma_R$ \\ 
   $\GeV^{-2}$       &$\GeV^{-2}$      &$\GeV^{-2}$      &                 & $\GeV^{-2}$       &$\GeV^{-2}$    & $\GeV^{-2}$  &           \\ \hline
  $15.701\pm0.001$&$4.323\pm0.001$&$24.709\pm0.002$&$7.819\pm0.0005~$&$0.2922\pm0.0005$&$1.540\pm 0.003$&$4.472\pm0.003$&$7.503\pm0.006$  \\ \hline 
\end{tabular*}
\end{table*}
}
 
{\small
\begin{table*}[ptb]
\caption{ $\chi^2$ (namely $\chi^2/d.o.f.)$ values for several ranges  of the 13 TeV data, with the 
analytical forms of the KFK model given by Eqs.(\ref{psi_st},\ref{hadronic_complete},\ref{R_tail}). 
  The {\it unique  solution}  given in  Table \ref{tableone}  is used in the  determination 
of $\chi^2$  for  all selected ranges shown in this table and in the plots of Fig.\ref{ranges}. 
   " With tail"  means that extra 3-gluon exchange  perturbative contribution   of Eq.(\ref{R_tail}) 
 is added to the analytical  basis.  
In the calculations of this table the CNI (Coulomb-nuclear interference) phase  is put equal to zero.
Values of $\chi^2$   calculated with statistical errors only and with total errors formed 
by quadrature of statistical and systematic errors are shown. The 39 points of 27.4 GeV enter with 
statistical errors only. }
  \label{tabletwo}                    
       \begin{tabular*}{\textwidth}{@{\extracolsep{\fill}}ccccccc@{}}
N  & $|t|$ range          &$\langle\chi^2\rangle$(total) &$\langle\chi^2\rangle$(total) & $\langle\chi^2\rangle$(stat) &   $\langle\chi^2\rangle$(stat)        & Remarks  \\ 
pts& $\GeV^{2} $          &  no tail  & with tail & no tail & with tail      &   \\ \hline 
138&0.000879-0.201041     &  0.0162   &  ---      & 1.455   &     & entire Set I   \\  
260& 0.000879-0.24902     &  0.2737   &   ---     & 4.852 &     & Set I(138) + Forward Set II (122) \\ \hline
122&0.0384-0.24902        &  0.5866   & ---       & 9.066   &    &   Forward part of Set II         \\  
144& 0.25643-0.89633      &  3.068    & ---       & 5.643 &     & dip-bump region in Set II       \\ 
152& 0.25643-1.15991      &  3.187    & ---       & 5.690  &      & extended dip-bump region in Set II  \\ 
24 & 0.91528-3.82873      &  10.74    & 12.72     & 11.73   & 13.70     & range of highest $|t|$ in Set II  \\  
290&0.0384-3.82873        &  2.326    & 2.448     & 7.052   & 7.164     & entire Set II        \\         
346&0.000879-3.82873      &  1.943    & 2.044     & 6.142   & 6.235    & Set I(56) + Set II(290) \\  
428&0.000879-3.82873      &  1.567    & 1.642     & 5.186  & 5.260    & Set I(138) + Set II(290) \\  \hline
385& 0.000879-14.2        &  ---      & 2.103     &---  & 5.869    & Set I(56) + Set II(290) + Faissler et al (39) \\  
467&0.000879-14.2         & ---       &  1.731    & ---  & 5.042     & Set I(138) + Set II(290) + Faissler et al (39) \\ \hline 
 \end{tabular*}
\end{table*}
 }
 {\small
\begin{table*}[ptb]
\caption{Quantities derived from the solution of the fitting of
  the  13 TeV data, with the parameters  given in Table \ref{tableone}. 
The quantities $Z_I$,  $Z_R^{(1)}$  and $Z_R^{(2)}$  are the locations ($|t|$ values) of the zeros 
of the imaginary and real amplitudes. Properties of the amplitudes are discussed in Sec.\ref{amplitudes}.  
  \label{tablethree}  }
    \begin{tabular}{@{\extracolsep{\fill}}ccccccccccc@{}}
      \hline
 \hline
   \multicolumn{3}{c} {Imaginary Amplitude} & \multicolumn{4}{c} {Real Amplitude}  & \multicolumn{1}{c} {Elast.} & \multicolumn{1}{c} {Inel.} &\multicolumn{2}{c}  {Dip}    \\ \cline{1-11}
  $\sigma$        & $Z_I$       &$B_I$          & $\rho$          & $Z_R^{(1)}$   & $Z_R^{(2)}$  &$B_R$          &$\sigma_{\rm el}$&$\sigma_{\rm inel}$ &$|t|_{\rm dip}$ &${\rm h_{dip}}$ \\ 
      mb          & $\GeV^{2}$  &$\GeV^{-2}$    & $          $    & $\GeV^{2}$    & $\GeV^{2}$   &$\GeV^{-2}$    &  mb        &  mb     & $\GeV^{2}$  & mb/$\GeV^2$       \\ \hline
  $111.56\pm0.01$ &$0.46\pm0.01$&$21.05\pm0.01$&$0.118\pm0.001$&$0.200\pm0.001$&$1.180\pm0.010$&$26.39\pm0.06$&$31.10$  &$80.46$ & 0.47 & 0.025  \\ \hline 
\end{tabular}
\end{table*}      }   

\subsection{Connection with measurements at $\sqrt{s}= 27.4  \GeV$ \label{Faissler} }

The elastic scattering data for   $|t|$ larger than 5 $\GeV^2$ have been shown 
to be independent of the energy in a large range of $\sqrt{s}$ from 20 GeV to 
7 TeV \cite{Faissler,Conetti,flavio1,KFK_1}. 
The experiment at $\sqrt{s}= 27.4  \GeV$ with 39 data points covering  the wide $|t|$  
range from   5.5  to 14.2 $\GeV^2$ \cite{Faissler}, provides  important 
reference for the study of  pp  at large scattering angles. 
 The property  is demonstrated for   energies below 
$\sqrt{s}$ = 62.5 GeV in   Fermilab and  CERN/ISR  measurements \cite{flavio1}
  (see Figures 2,3 and  10 in this paper), 
showing  a smooth connection between the mid-$|t|$   range 
containing  perturbative and nonperturbative terms  and the range 
of large $|t|$  dominated by perturbative three-gluon exchange. 
 
The universality in the energy  and the $|t|$
dependence of  form  $1/|t|^8$  in  $d\sigma/dt$ have been interpreted  
by Donnachie and Landshoff \cite{DL1}  as determined by 
the process of exchange of three gluons.  
This contribution is represented by the quantity $R_{ggg}(|t|)$ introduced 
in Eq.(\ref{hadronic_complete}), receiving a cut-off factor written in 
Eq.(\ref{R_tail}) designed to restrict   the $1/|t|^8$ behaviour. 
The three-gluon contribution occurs  in the $|t|$ range where the imaginary 
part is negligible, and the  perturbative term  $\alpha_R ~ \exp({-\beta_R |t|})$  
 is dominant.  The transition from  2 to 5 $\GeV^2$ is precious to inform 
features (signs, magnitudes) of terms of the real scattering amplitude in 
the large $|t|$ region. These  features are described in Sec.\ref{amplitudes}.

 As an example, the structure of the real  amplitude  leads to the 
argument that the difference in the dip regions  of pp and ${\rm p \bar p }$  
 scattering at 53 GeV  \cite{Breakstone} is due to the difference 
in the signs of the three-gluon contributions  in pp and ${\rm p \bar p }$  
scattering, and not necessarily to the presence of an odderon element \cite{flavio1}, 
unless it is meant that three-gluon exchange is the modern QCD name for odderon
\cite{Ewerz,Levin}. 
 
At high energies, there is not  sufficient  experimental information for the    
investigation  of the elastic amplitudes at high $|t|$. LHC  measurements  at 
7 and 8 TeV \cite{KFK_2,KFK_3} are restricted to less than  $|t| = 2 ~ \GeV^2$, 
and the connection between mid and large $|t|$ regions remains in the level of 
{\it clear indication}, as shown in Fig.6 of the 7 TeV paper  \cite{KFK_2},
where the data for 52.8 GeV \cite{Nagy,Amos} and 7 TeV  are exhibited together.
 
\begin{figure*}[b]
 \includegraphics[width=8cm]{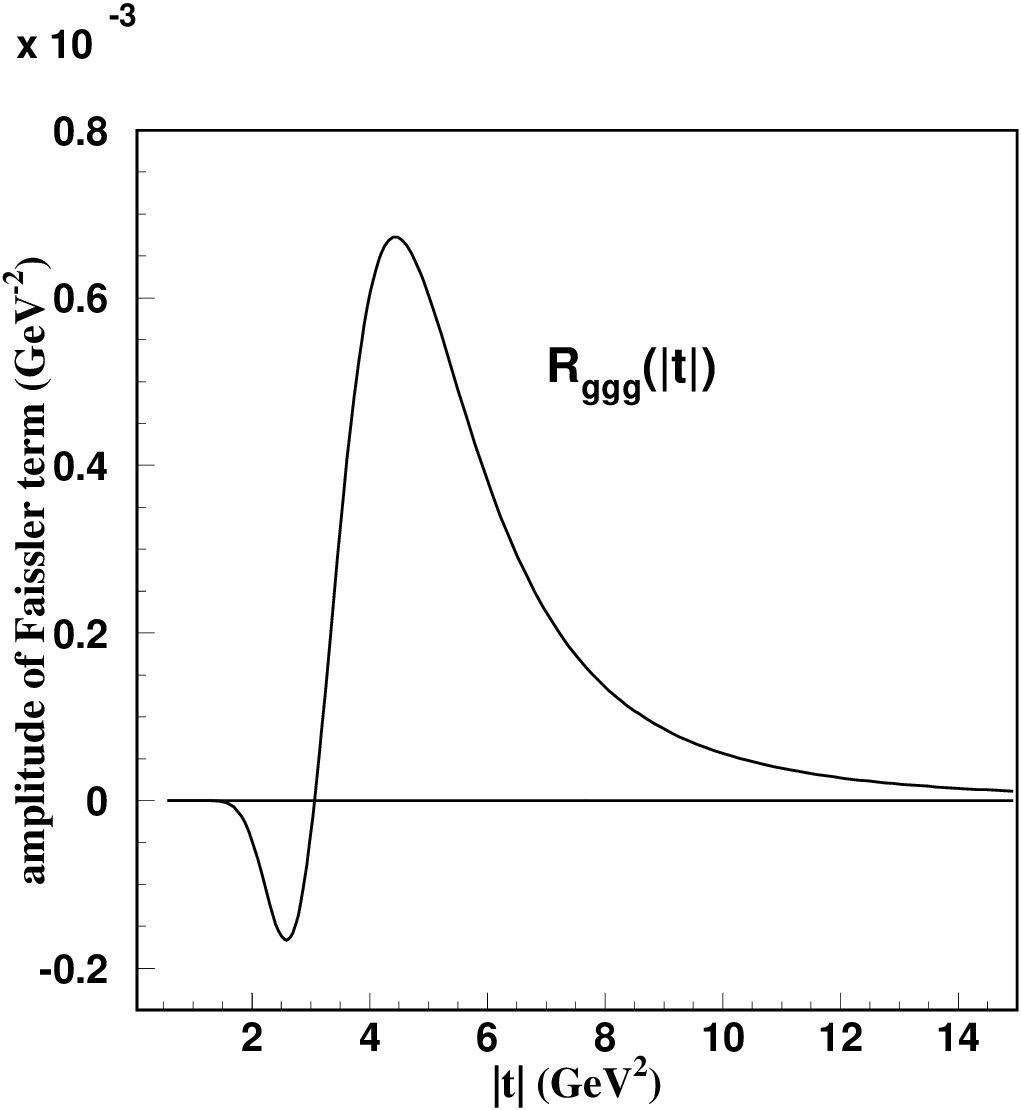} 
 \includegraphics[width=8cm]{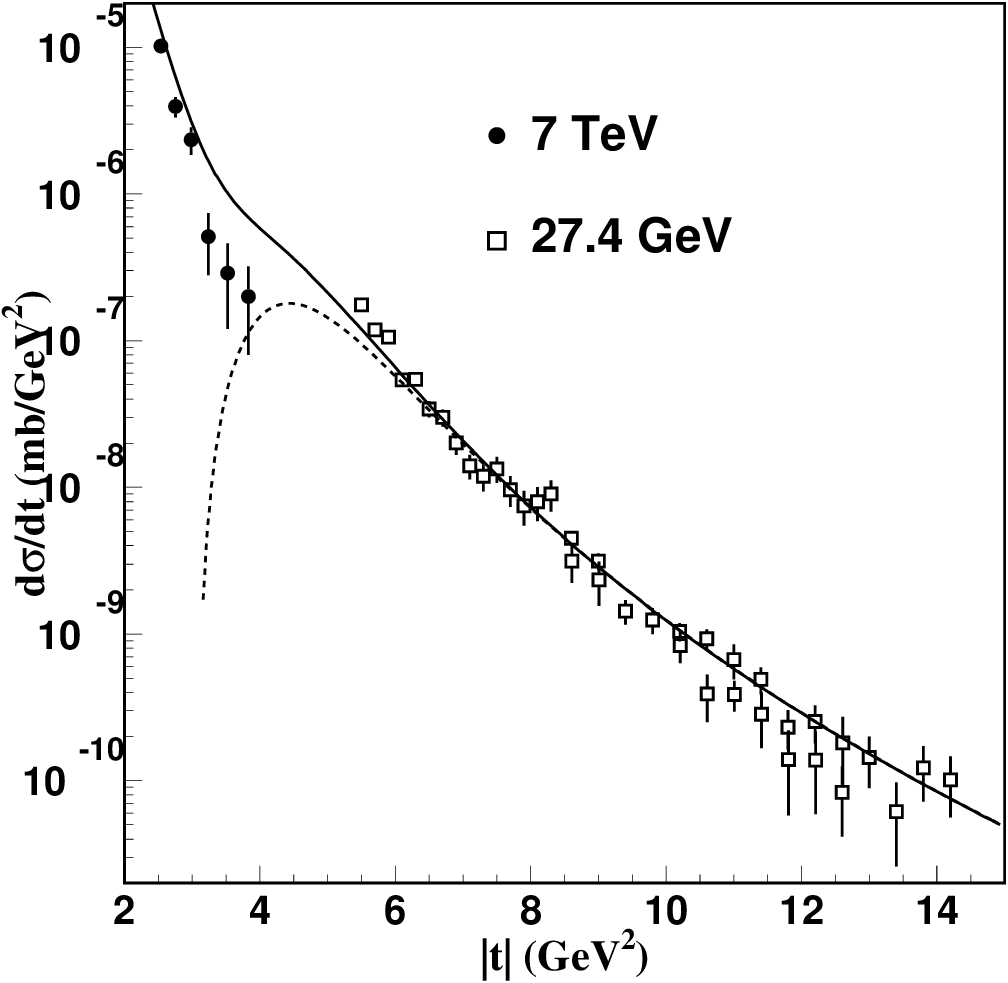} 
 \caption{  \label{connection} Connection of the low energy ($\sqrt{s}=27.4 \GeV$) points of  large $|t|$   
($ 5.5\leq |t||\leq 14.2 ~\GeV^2$)    with Totem 13 TeV data.
a) Form proposed for the amplitude $R_{ggg}(|t|)$ in Eq.(\ref{R_tail}) for three-gluon exchange
with a cut-off factor acting for $|t|\leq 4 \GeV^2$.
b) Differential cross section calculated including the  $ R_{ggg}(|t|)$ term
(solid line) plotted together with   points of Totem measurements at 13 TeV 
(full circles) and the points (open squares) at 27.4 GeV. The piece of dashed line
pointing downwards shows the action of the cut-off factor.  }
\end{figure*}

\begin{figure*}[b]
 \includegraphics[width=8cm]{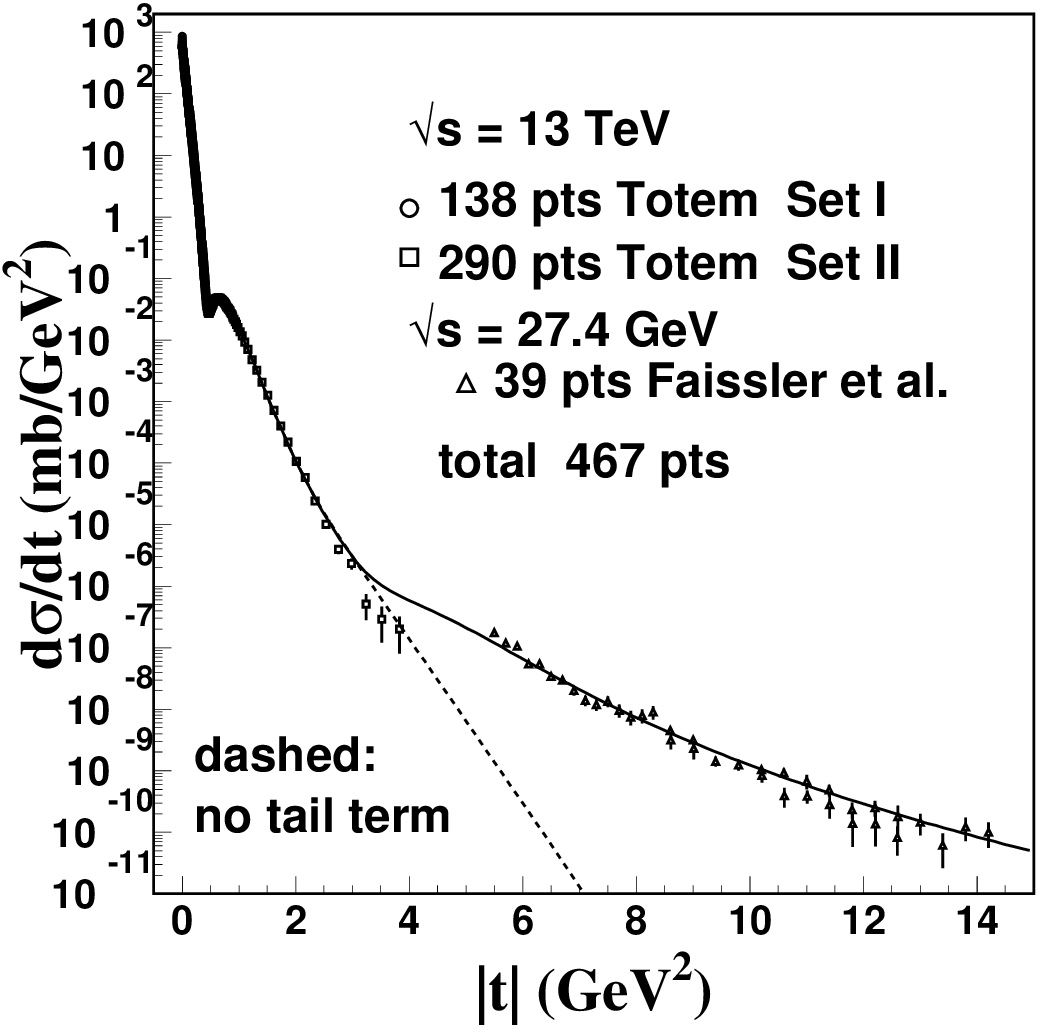} 
  \includegraphics[width=8cm]{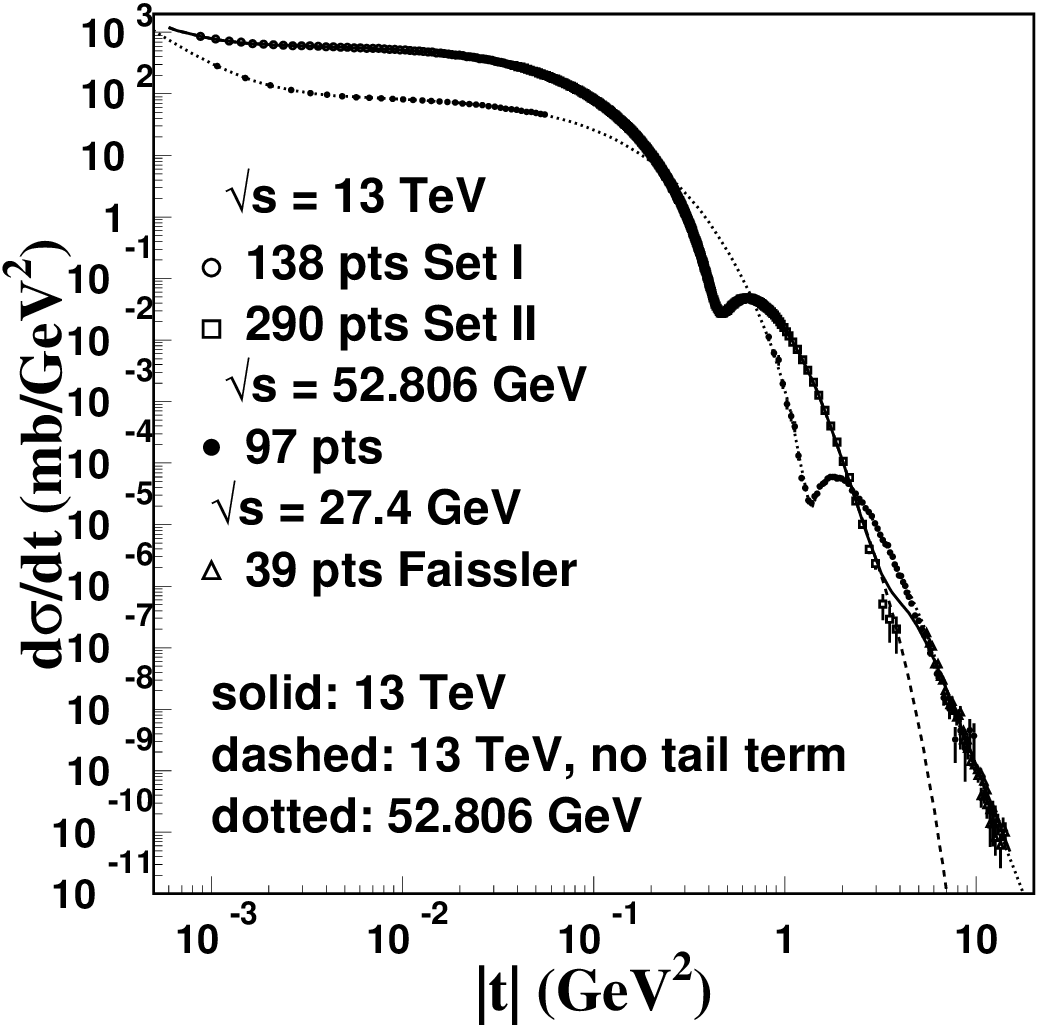} 
\caption{ a)  Analytical representation for all 467 data  points : 138 points of Set I \cite{Totem_13_1} 
 and 290 of Set II \cite{Totem_13_2}   from  Totem measurements  at 13 TeV, 
plus 39 points at $\sqrt{s}$= 27.4 GeV from FNAL  measurements \cite{Faissler}.   
The global representation  of 467 (138+290+39) points   is constructed  
with the unique solution given in Table \ref{tableone} plus the $R_{ggg}(t)$ term as in 
Eq.(\ref{hadronic_complete}), with   results $\chi^2=1.731$ (total statistical and systematic errors)
and $\chi^2=5.042$ (statistical errors), as shown in Table \ref{tabletwo}. 
The dashed lines represent the analytical form for 13 TeV  excluding the 3-gluon exchange tail term. 
b) Joint plot of data at 52.806 GeV \cite{Nagy,Amos} and Totem 13 TeV data, with the analytic
solutions obtained with the KFK model \cite{KFK_2}. The points have energy scales differing by 
more than 200, and still the data in the large $|t|$ region  have similar magnitudes. 
The universality is consistently present at  ISR energies \cite{flavio1,flavio2}.  
\label{NOVAS} }
\end{figure*}

At 13 TeV the data are more extended in  $|t|$, reaching  nearly 4 $\GeV^2$, 
allowing investigation of properties of the amplitudes in the connection with FNAL 
data \cite{Faissler}. Then we first choose the parameters for the $R_{ggg}(t)$ 
function, that is shown Fig.\ref{connection}, together with the corresponding 
cross section in the range of the transition. 
In Fig.\ref{NOVAS} we show the matching of the Totem 13 TeV 
and  ISR 52.806 GeV measurements \cite{Nagy,Amos}      with 
the data of  FNAL  measurements \cite{Faissler} at $\sqrt{s}=27.4 \GeV$. 



Some  points of high  $|t|$ of the Totem measurements  show  a marked decrease in the values
of $d\sigma/dt$, with large statistical error bars, from 45\% to 60\%.  These points deviate 
meaningfully from the proposed solution, and particularly they seem not to accept easily the suggestion 
of connectivity with the three-gluon tail. These are only few points of poor statistics, but visually 
they have important influence, as shown in Figs.\ref{connection} and \ref{NOVAS}. In our description,
this range  of $d\sigma/dt$ is  dominated by the perturbative term in the real amplitude, and 
serves as important test of the proposed disentanglement. In Sec.\ref{amplitudes} we show that
the real part of the KFK  amplitude is  positive for large $|t|$, and then the superposition with the 
also positive three-gluon term should be constructive. If the real part were negative, a dip 
could be formed. In the analysis of the 1.8/1.96 GeV Fermilab \cite{KFK_1} $\rm{p \bar p}$ data 
 we predicted that such dip would appear for  large $|t|$ (the three gluon term is negative 
in   $\rm{p \bar p}$ ), but unfortunately the measurements  do not reach large enough $|t|$, 
and the prediction is not tested. Here in pp  at 13 TeV, we do not have simple explanation for the 
decrease of $d\sigma/dt$ in the points of largest $|t|$. A connection function producing the
visual shape would not be natural. This question obviously  leads to the   suggestion that the  
measurements in the large $|t|$ range should receive more attention. 

   Table \ref{tabletwo} shows that the 24 points of with highest $|t|$ in Set II  are described 
in our unique solution with comparatively large $\chi^2$  values of about 10. This is 
a local feature, as these points have low influence in the $\chi^2$ value for the 428 points.    
For a local investigation, we  observe that  this range is dominated by the perturbative real part, 
so that only the parameters $\alpha_R$ and $\beta_R$  require attention. Thus, with 
$\alpha_R=0.476\pm0.022$ and $\beta_R=1.771\pm0.025$ we obtain $\chi^2=2.210$ and $\chi^2=2.484$, 
respectively using total and only statistical errors.  This predicted local improvement in 
$\chi^2$ changing only two selected parameters is consequence of the separation of the  
perturbative and non-perturbative terms in the analytical form. 

\subsection{ \label{forward} Specific representation of amplitudes  for the 138 points of Set I  }

As a side information (since the main concern of the present work is with the unique global solution 
for all ranges), in Table \ref{tablefour} we show the $\chi^2$  results for the 138 points of Set I 
with freedom given to the $\lambda_I$ and  $\lambda_R$  
parameters,   maintaining all other quantities as written and used in Tables \ref{tableone}
and   \ref{tabletwo}. Only the non-perturbative magnitudes $\lambda_I$  and  $\lambda_R$  are 
investigated in this alternative examination because these terms are dominant in the imaginary and 
real amplitudes for small $|t|$, as shown in Sec.\ref{amplitudes}. 
  Comparison is made of solutions with and without inclusion of the Coulomb-nuclear 
interference phase  $\phi$.
The results  in Table \ref{tablefour}  may be compared with values   obtained with simplified forms 
of amplitudes restricted to the forward 
scattering range  \cite{PLB2019}, namely with product of exponential and linear factors  as 
\begin{equation}
     T_K(t) = T_K(0) ~  e^{(B^0_K/2) t}(1-\mu_K ~t)  ~ ~ ~ , K=I,R ~ .  
\label{forward_model}
\end{equation}
 We again  stress that the parameters $\sigma$, $\rho$ and slopes are model-dependent quantities, related 
to specific analytical forms of the amplitudes. The only experimental measurements are 
the values of  $d\sigma/dt$ at angular positions defined by values of $|t|$.
  In particular, for 
the   value of   $\rho$, it has been shown   \cite{PLB2019} that the presently available  
data at small $|t|$ does not allow a conclusion about its value. Besides the insufficiency of 
regular data in the very forward region, the  theoretical basis for the Coulomb-nuclear 
interference phase is uncertain. 

We remark that both imaginary and real parts have zeros, so that, besides exponential slopes 
at least linear factors in the amplitudes  are essential to represent the forward  data  realistically
as in Eq.(\ref{forward_model}). 
 In KFK   the   factorization of the logarithmic derivative with a slope  as in  Eq.(\ref{slopes_par})   
leaves a remainder that has a zero, but not a linear zero (actually the remaining factor has 
zero of higher order in a Taylor expansion), so that  $B_I$ and $B_R$ in Eq.(\ref{slopes_par})
correspond to the {\it effective} slope that includes the effect of a linear factor in the forward 
amplitude. The effective slope in Eq.(\ref{forward_model})  comparable to Eq.(\ref{slopes_par})  is 
      $B^{\rm eff}_K = B^0_K - 2 \mu_K$.  
It is also interesting to compare the value of the first real zero $Z_R^{(1)}$ of the KFK model in
Table \ref{tablethree} with the values obtained  \cite{PLB2019} with Eq.(\ref {forward_model}). 
With $\mu_R=-3.84 ~\GeV^{-2}$ , the zero at $|t|=-t=-1/\mu_R=0.26 \GeV^2$  may be compared with 
$Z_R=0.20 \GeV^2$ in Table \ref{tablethree}. 
  {\small
\begin{table*}[ptb]
\centering
\caption{Values of parameters $\lambda_I$ and $\lambda_R$
and of  $\chi^2$ (with statistical errors only)  obtained  specifically   for the 138 points of Set I, 
 with all  other quantities ($\alpha_K,~\beta_K,~\gamma_K$)  kept as  given in Table \ref{tableone} 
and used in Table \ref{tabletwo}. We here give the values for fitting with Coulomb interference 
phase $\phi$  put as zero, and for phase calculated as described 
before \cite{PLB2019}.  The $\chi^2$ values 
may be compared with $\chi^2=1.455$ (with CNI phase zero) given in Table \ref{tabletwo} and $\chi^2=2.144$ 
with   CNI phase   calculated with proton form factor. 
We recall that in the detailed analysis of forward data studying the influence of the CNI phase \cite{PLB2019},
  reported values are $\sigma=111.84 ~{\rm mb},~ \rho=0.125$ for $\phi=0$ , and $\sigma=111.84 ~ {\rm mb}, ~
\rho=0.097$ for $\phi \neq 0$. The position $Z_R^{(1)}$ of the first real zero (Martin's Zero) 
  is also given, since it occurs in the forward range and is important theoretical reference.  
      \label{tablefour}    } 
\begin{tabular}{@{\extracolsep{\fill}}ccccccccc@{}}
      \hline
 \hline
CNI          & $\lambda_I$    &  $\lambda_R$    &$\chi^2$& $ \sigma$     & $\rho$        & $B_I$     & $B_R$      & $Z_R^{(1)}$   \\  
phase $\phi$ &  $\GeV^{-2}$   &  $\GeV^{-2}$    &        &      mb       &               &$\GeV^{-2}$&$\GeV^{-2}$ &  $\GeV^2$ \\ \hline
 zero        &$24.772\pm0.010$&$4.382\pm0.115$&1.126   &$111.73\pm0.03$&$0.116\pm0.001$& 21.06      & 26.37      &  0.201  \\ \hline 
$\phi(t)$    &$24.836\pm0.010$&$3.403\pm0.130$&1.121   &$111.91\pm0.03$&$0.092\pm0.001$& 21.08      & 25.96     & 0.213    \\ \hline
 \end{tabular}
\end{table*}
}  


\section{ \label{amplitudes} Imaginary and real parts of the scattering amplitude } 

The  analysis presented in Sec.\ref{data_analysis} leads to  a proposal  for the disentanglement of 
the real and imaginary parts, that is obtained directly from the data. In this section we discuss 
the properties of the amplitudes and their terms, in both $t$ and $b$ coordinates.

\subsection{Amplitudes  in $t$ space      \label{t-amplitudes}    }  

\begin{figure*}[b]
     \includegraphics[width=8cm]{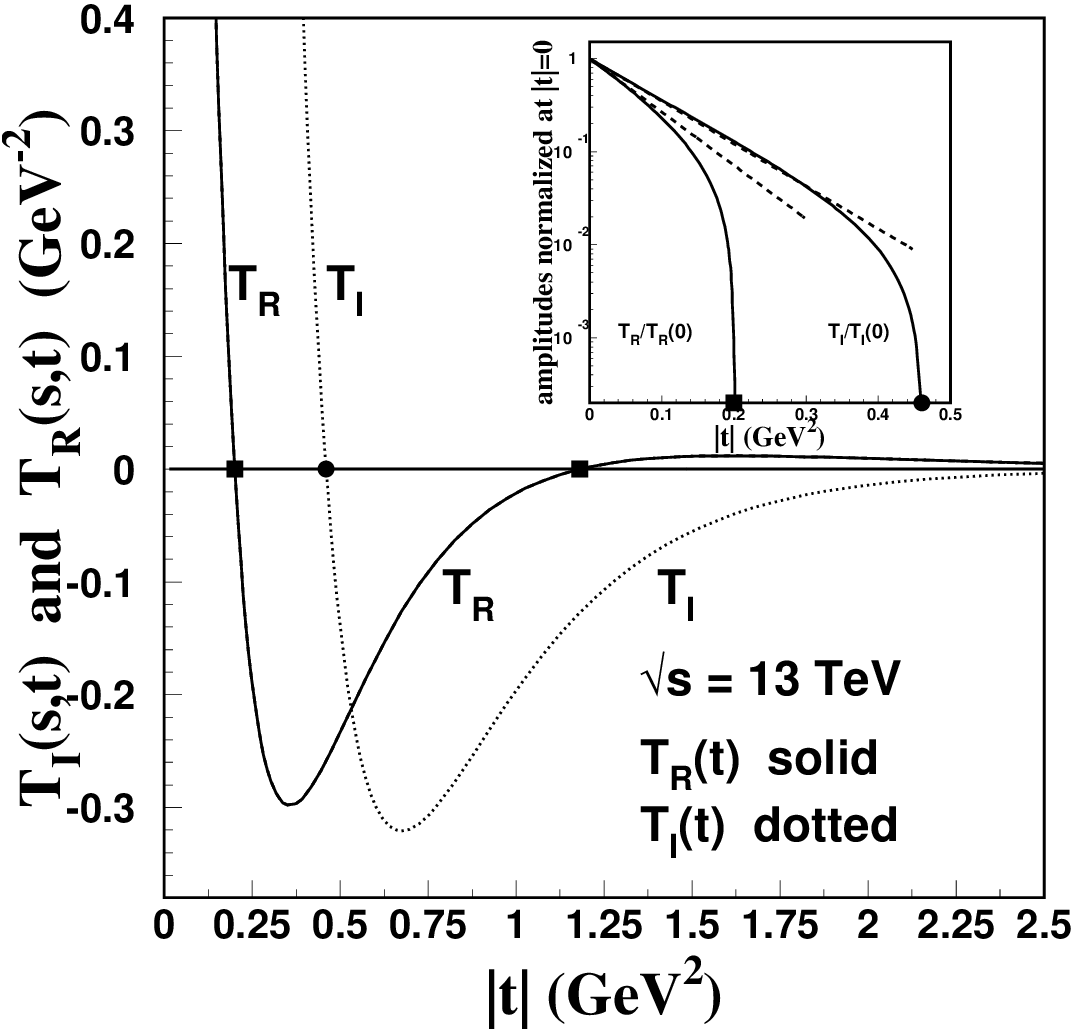} 
     \includegraphics[width=8cm]{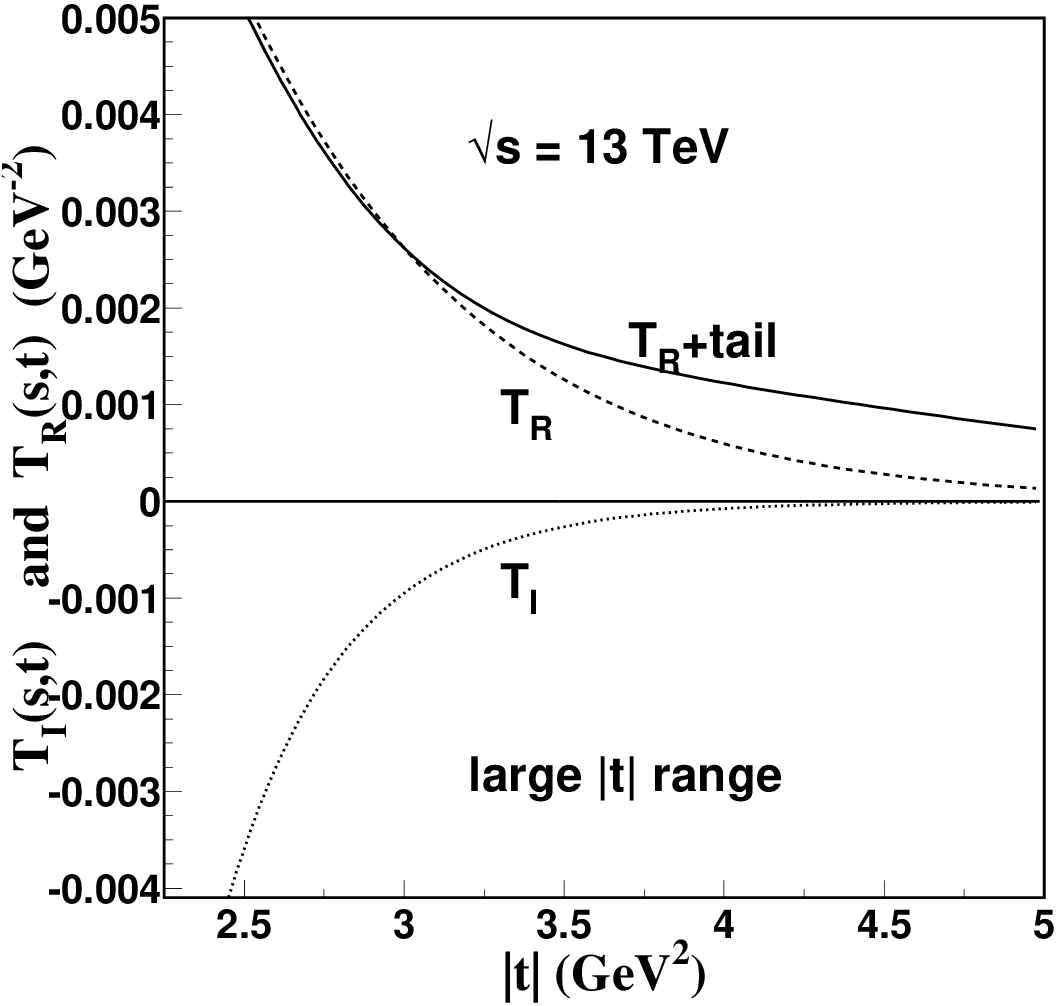} 
\caption {$|t|$ dependence of the real and imaginary parts of the pp  elastic amplitude
at $\sqrt{s}$ = 13 TeV, showing one zero at $Z_I=0.46~ \GeV^2$ for $T_I(t)$, and  zeros
at $Z_R^{(1)}=0.200~ \GeV^2$ and $Z_R^{(2)}=1.180 ~\GeV^2$ for $T_R(t)$. 
The inset uses log scale to exhibit  the slopes at $|t|=0$, demonstrating  the early deviation 
of the amplitudes from the linear behaviour, each amplitude bending towards its zero.  
For large $|t|$ the negative imaginary amplitude (dotted line) becomes negligible, and there is
strong dominance (in magnitude) by the positive real part (dashed line). 
   For  $|t|$ above $\sim  3 \GeV^2$ the three-gluon exchange contribution added to 
the real part (solid line)  raises  $d\sigma/dt$, forcing the behaviour observed at $\sqrt{s}=27.4 \GeV$  
and conjectured to be universal. The continuity in the inclusion of the three-gluon exchange term 
 is shown in Fig.\ref{NOVAS}.  
 \label{amplitudes_1-fig} } 
\end{figure*}

  Fig.\ref{amplitudes_1-fig} shows the amplitudes, detailing 
    small and large $|t|$ ranges. Similarly to lower energies,
the imaginary and real parts have one and two zeros respectively. In the 
plot for large $|t|$, the contribution of the the $R_{ggg}$ tail term is also shown, 
appearing as a deviation in the real amplitude visible for $|t| \geq 3  \GeV^2$.   

\begin{figure*}[b]
                     \includegraphics[width=8cm]{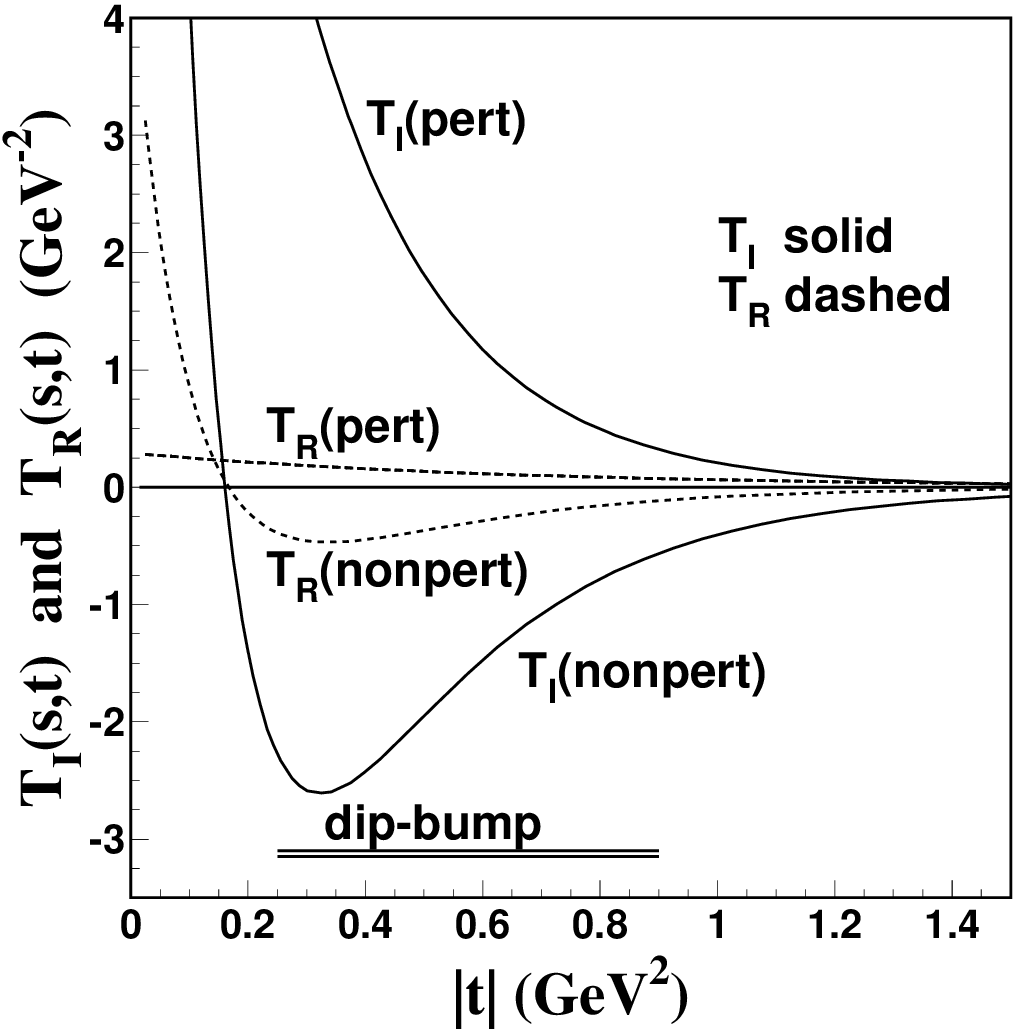}          
                     \includegraphics[width=8cm]{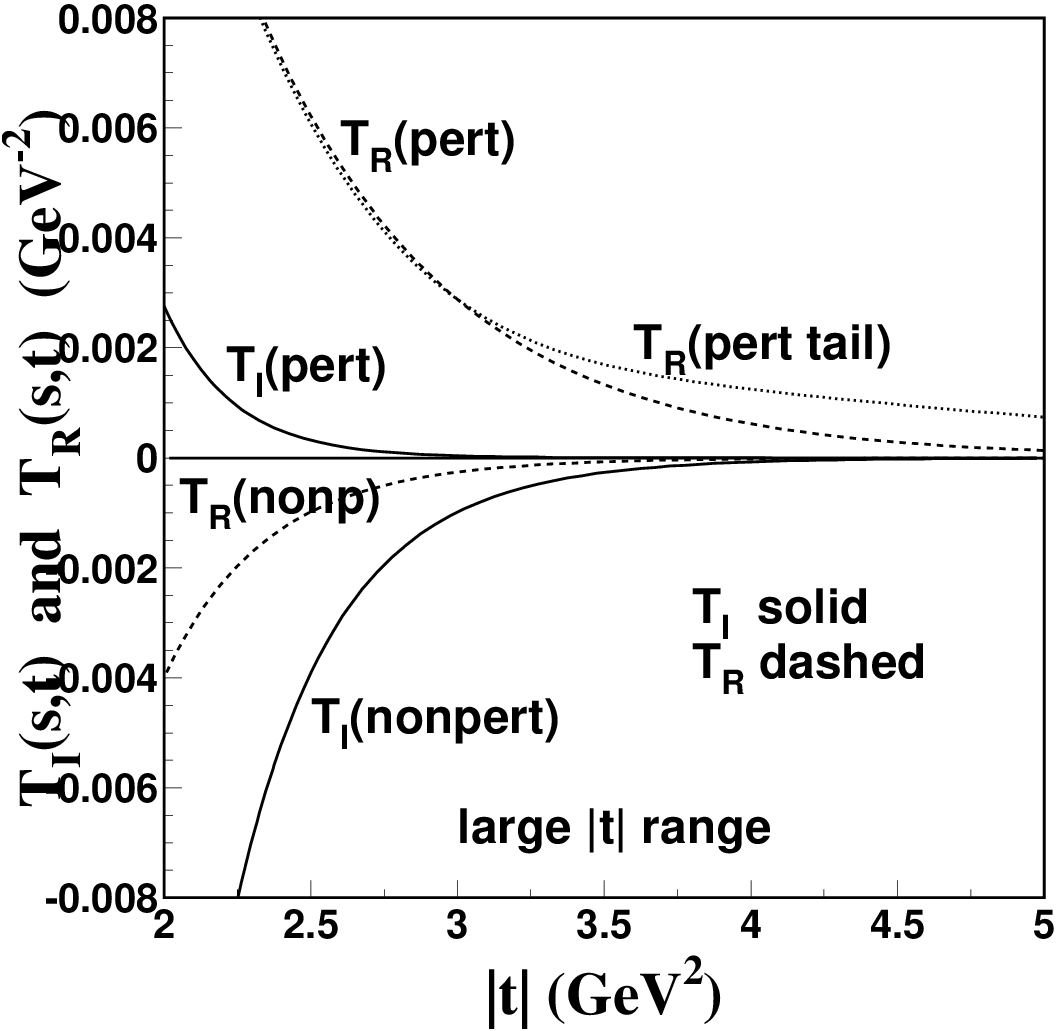} 
\caption { Perturbative and non-perturbative contributions in $T_I(t)$ and $T_R(t)$. 
In the figure we call $T_K({\rm pert}) = \alpha _{K} \mathrm{e}^{-\beta_{K} |t|}$  ,  
$T_K({\rm nonpert})=  \lambda _{K} \psi_{K}(\gamma _{K} ,t) $, (with $K=I,R$) ,  
  and $T_R ({\rm pert ~ tail})  = T_R({\rm pert})+ R_{ggg}(t)$.  
It is important to observe that $\lambda_I/\alpha_I \approx 25/15 $ and 
$\lambda_R/\alpha_R \approx 15 $, so that the forward direction is dominated by   
the non-perturbative term, particularly so in the real amplitude (thus the 
evaluation of the $\rho$ 
parameter is mainly a non-perturbative affair).  After the bump, $T_I({\rm pert})$  is 
negligible compared to $T_I({\rm nonpert})$, which  becomes negligible compared 
to $T_R({\rm pert})$ for $|t|\geq ~ 3 ~\GeV^2$. For large $|t|$, only 
$T_R({\rm pert})$ (or  $T_R({\rm pert ~ tail})$ ) survives.
 \label{amplitudes_2-fig} }
\end{figure*}

The separate  perturbative and nonperturbative parts of the imaginary and real amplitudes
 are shown in Fig.\ref{amplitudes_2-fig}. The quantities.
$T_I({\rm pert}) = \alpha _{I} \mathrm{e}^{-\beta_{I} |t|}$ 
and $T_I({\rm nonpert})=  \lambda _{I} \psi_{I}(\gamma _{I},t) $
are strong and with opposite signs in the dip-bump region, with a cancellation at 
$Z_I=0.46 ~\GeV^2$, causing the dip. The existence of these two terms in $T_I$ is 
most important for the construction  of the representation. The cancellation 
leaves room for the influence of the real amplitude that modulates the shape of
the dip-bump structure.   
  $T_R({\rm nonpert})$   dominates (in magnitude) over $T_R({\rm pert})$ in the 
dip-bump region, but if falls 
to zero more rapidly, while  the perturbative real part lasts longer in $|t|$.  
For $|t|$ larger than $\sim 3 ~\GeV^2$  only the perturbative real part $T_R({\rm pert})(t)$ 
remains active, with positive sign.

As a general view, we observe that forward scattering emphasizes
 non-perturbative dynamics, while large $|t|$ 
scattering is dominated by perturbative terms  in the  real amplitude. 
The real part becomes negligible for $|t|=0$, as $\rho$ decreases 
  with the energy.   

 The  magnitudes of all terms in 
the amplitudes  vary enormously from the bump to the region $|t|$ = 
3-4 $\GeV^2$  reached by the  present data.  The
structure in the large $|t|$ range  that we try to access through the connection 
with the three-gluon exchange is  important  for the  construction of 
a global picture for pp  elastic scattering. This construction  
  is confirmed by other models, as illustrated in Fig.(\ref{BSW_SEL}).

\subsection{  \label{b_amplitudes} Amplitudes in $b$-space }   

\begin{figure*}[b]  
                  \includegraphics[width=8cm]{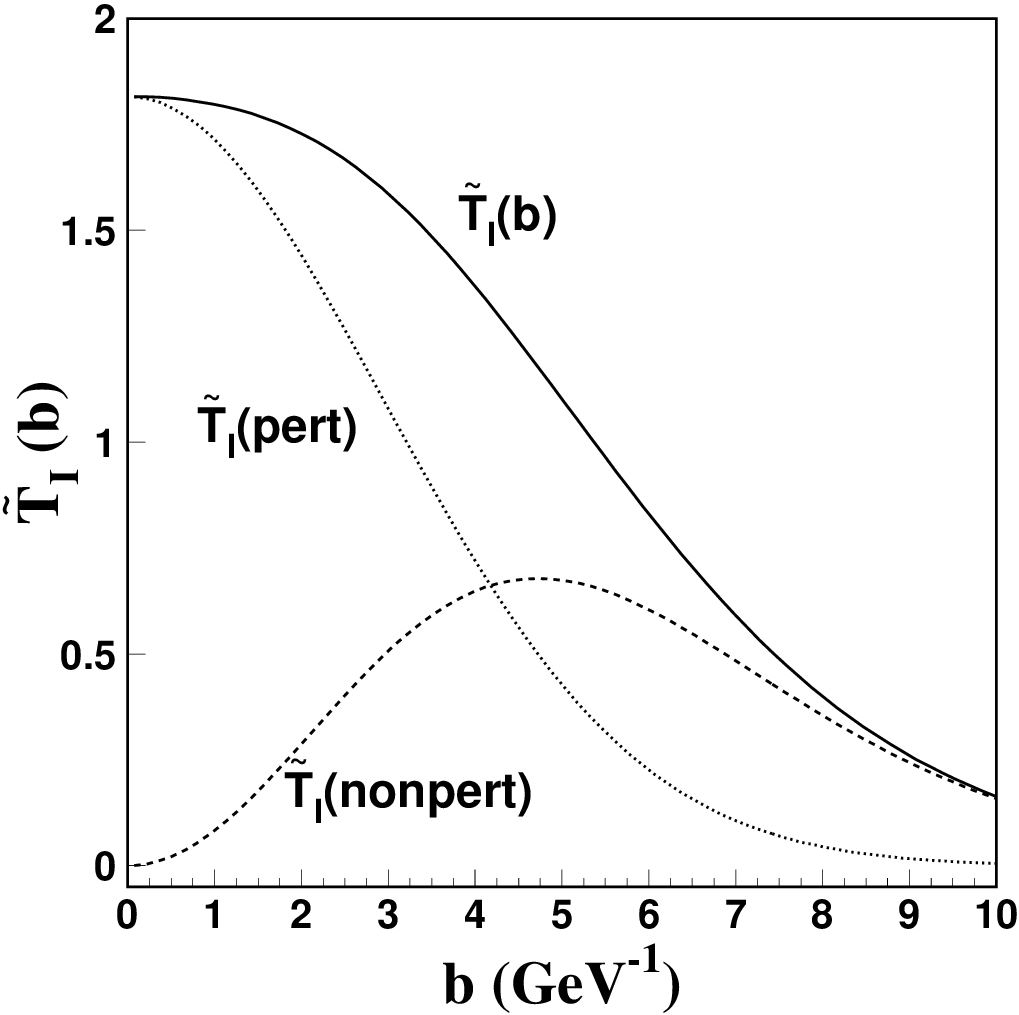} 
                  \includegraphics[width=8cm]{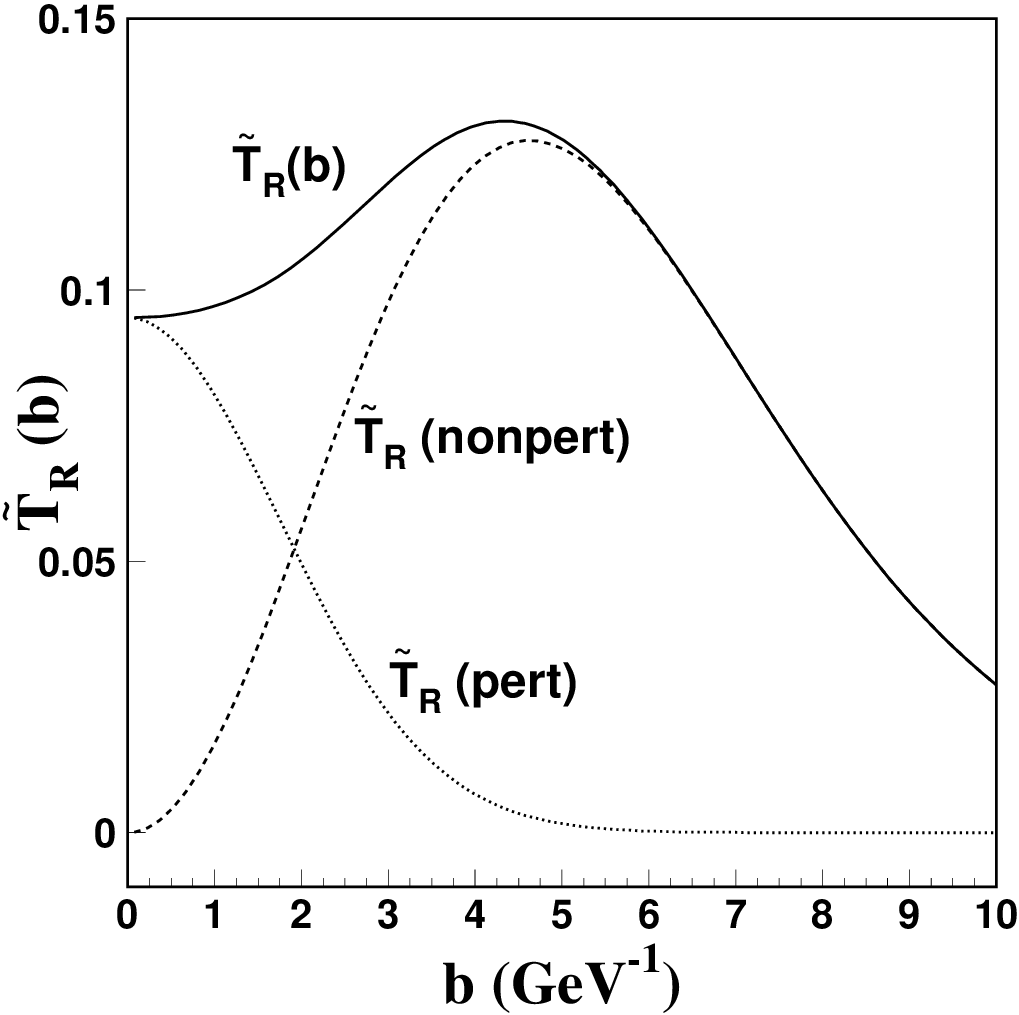} 
                   \includegraphics[width=8cm]{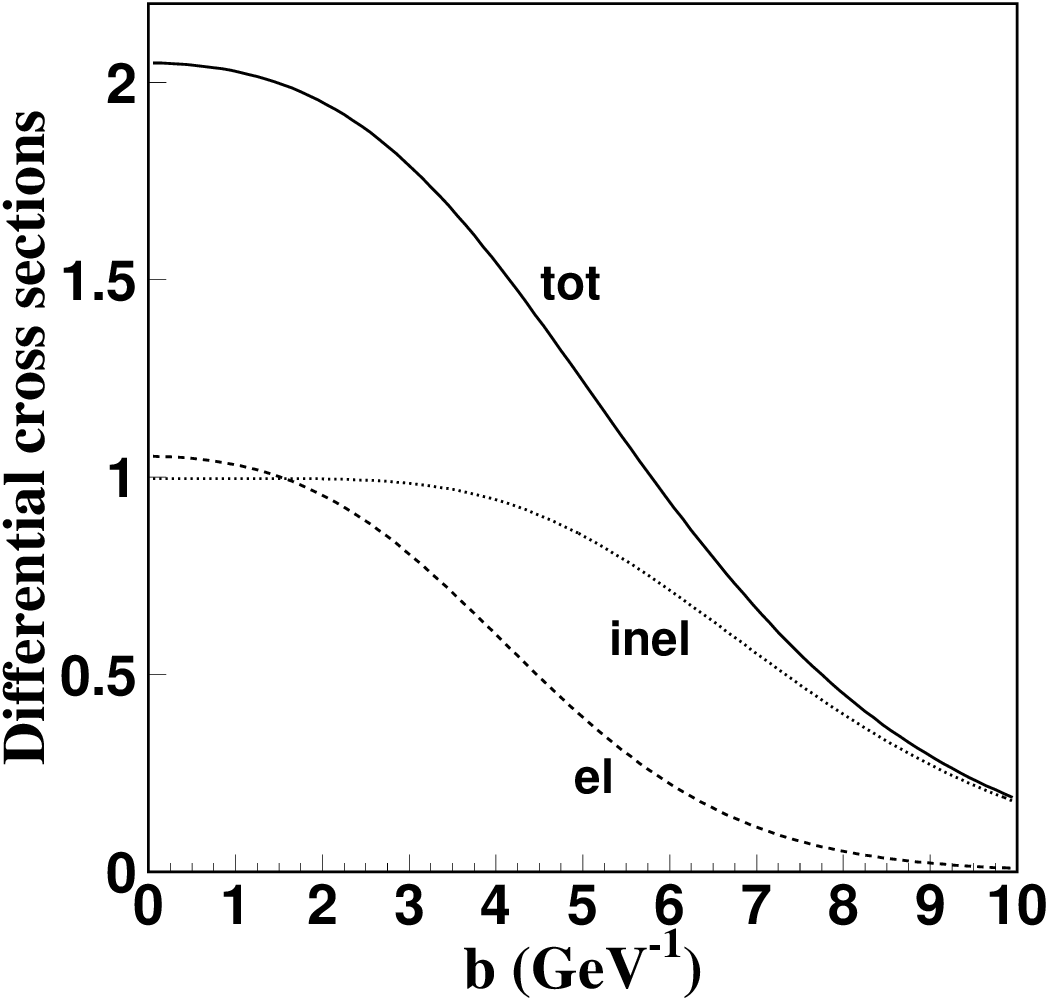} 
\caption {a),b)-  Amplitudes in b-space.   
 \label{b-amplitudes} 
The quantities labelled in the figures are 
$\widetilde{T}_{K}({\rm pert})=({\alpha _{K}}/{2\beta _{K}})e^{-{b^{2}}/{4\beta _{K}}}  $ ,  
 $\widetilde{T}_{K}({\rm nonpert})= \lambda _{K}\widetilde{\psi }_{K}(b) $,  
with $\widetilde{\psi }_{K}(b)$ given in Eq.(\ref{Shape-b}),
and $\widetilde{T}_{K}= \widetilde{T}_{K}({\rm pert})+\widetilde{T}_{K}({\rm nonpert})$.
Notice the difference in the scales of the plots of $\widetilde{T}_{I}$ and $\widetilde{T}_{R}$. 
The perturbative  terms dominate the central region of $b\leq 2 ~\GeV^{-1}\sim 0.4 ~ {\rm fm} $  
while the non-perturbative 
terms are strongly dominating for large $b$. In c) the plots of 
differential cross sections of Eqs.(\ref{dsdb_el},\ref{dsdb_sigma},\ref{dsdb_inel})
give hints about the structure of the interaction as observed in the transverse collision plane. 
 }
\end{figure*}

The $b$-space dimensionless amplitudes $\widetilde{T}_{I}(b)$  and $\widetilde{T}_{R}(b)$ 
of Eqs.(\ref{b-AmplitudeN},\ref{Shape-b}) are shown in Fig.\ref{b-amplitudes}a,b,
 where we observe that there are no zeros.    
In general $\widetilde{T}_{I}(b)$ is about 10 times larger than $\widetilde{T}_{R}(b)$,
and it is impressive that the Fourier  transforms of both have 
importance in  the structure of the observed $d\sigma/dt$, with 
a dominance of the real part for large $|t|$.  
The function $\widetilde{T}_{I}(b)$ is monotonically decreasing in $b$, 
while  $\widetilde{T}_{R}(b)$ has a maximum at $b=4.339 ~\GeV^{-1}$
with numerical value 0.131.
At $b=0$ we have 
$$  \widetilde{T}_{I}(b=0)= \alpha_I/2 \beta_I =  1.81598 = \sqrt{\pi}+0.04353   $$
that is slightly  larger than $\sqrt{\pi}=1.7725$ 
and 
$$  \widetilde{T}_{R}(b=0) =\alpha_R/2 \beta_R  = 0.09487 ~ .  $$
At 
$$ b=b_{\rm root}= 1.47393 \GeV^{-1} $$
we have 
$$ \widetilde{T}_{I}(b_{\rm root} )=\sqrt{\pi}=1.7725  ~ ~\   {\rm and} ~ ~\    
   \widetilde{T}_{R}(b_{\rm root})= 0.10009 ~ .   $$

As seen in Fig.\ref{b-amplitudes}, the non-perturbative terms 
$\widetilde{T}_{K}({\rm nonpert})= \lambda _{K}\widetilde{\psi }_{K}(b)$, $K=I,R$, dominate
the amplitudes for large $b$, while  $\widetilde{T}_{I}({\rm nonpert})$ 
dominates over $T_I(t)$ in the forward peak, where non-perturbative and perturbative 
magnitudes are in the ratio  $\lambda_I/\alpha_I \sim 25/15$, with a ratio 
$\sim 25/9$  in the  contributions to the total cross section. 
It is  remarkable that forward elastic  scattering  
  is mainly a peripheral process of   non-perturbative nature. 

In terms of the $\widetilde{T}_{K}(s,\vec{b})$ amplitudes, the elastic,
total and inelastic cross sections are written respectively  
\begin{equation}
\sigma _{\mathrm{el}}(s) =\frac{(\hbar c)^{2}}{\pi }\int d^{2}\vec{b}~|%
\widetilde{T}(s,\vec{b})|^{2} \equiv \int d^{2}\vec{b}~\frac{d\widetilde{%
\sigma }_{\mathrm{el}}(s,\vec{b})}{d^{2}\vec{b}}~,
\label{dsdb_el}
\end{equation}%
\begin{equation}
\sigma(s) =\frac{2}{\sqrt{\pi }}(\hbar c)^{2}\int d^{2}\vec{b}~\widetilde{T}%
_{I}(s,\vec{b})~ \equiv \int d^{2}\vec{b}~\frac{d\widetilde{\sigma }_{%
\mathrm{tot}}(s,\vec{b})}{d^{2}\vec{b}}~~,
\label{dsdb_sigma}
\end{equation}
and 
\begin{eqnarray}
\sigma _{\mathrm{inel}} &=&\sigma-\sigma _{\mathrm{el}} =(\hbar c)^{2}\int
d^{2}\vec{b}~\Bigg(\frac{2}{\sqrt{\pi }}\widetilde{T}_{I}(s,\vec{b})-\frac{1%
}{\pi }|\widetilde{T}(s,\vec{b})|^{2}\Bigg)  \notag \\
&\equiv &\int d^{2}\vec{b}~\frac{d\widetilde{\sigma }_{\mathrm{inel}}(s,\vec{%
b})}{d^{2}\vec{b}}~.  \label{dsdb_inel}
\end{eqnarray}%
The values of the integrated cross sections are 
$\sigma_{\rm el} = 31.096 $ mb, $\sigma  = 111.557 $ mb, 
 $\sigma_{\rm inel} = 80.461  $ mb, 
with  ratio  $\sigma_{\rm el}/\sigma=0.28$ .    
The differential cross sections in $b$-space shown in Fig.\ref{b-amplitudes}c
give a hint of the proton hadronic interaction structure in the transverse collision
plane with smooth monotonous $b$-dependence.

Unitarity imposes that $\sigma_{\rm el} \leq \sigma$.  With a classical point of view, a 
hypothesis that the inequality is  valid for all $b$ is written
\begin{equation}
\widetilde{T}_{I}(s,\vec{b})^2+\widetilde{T}_{R}(s,\vec{b})^2 \leq 2 \sqrt{\pi}~  \widetilde{T}_{I}(s,\vec{b}) ~ , ~\forall ~s,b~  
\label{b-unitarity_0}
\end{equation}  
or
\begin{equation}
\widetilde{T}_{R}(s,\vec{b})^2+(\widetilde{T}_{I}(s,\vec{b})-\sqrt{\pi})^2 \leq  \pi   ~  , ~ ~\forall ~s,b~  . 
\label{b-unitarity} 
\end{equation}
This relation, called $b$-space unitarity, is satisfied by our amplitudes.

The  eikonal function $\chi \left( s,b\right) $ for a given $s$   is introduced  through 
\begin{equation}
i\sqrt{\pi }~(1-e^{i\chi (b)})~\equiv \widetilde{T}(b)=%
\widetilde{T}_{R}(b)+i\widetilde{T}_{I}(b),  \label{Eikonal}
\end{equation}%
with 
\begin{equation}
\chi (b)=\chi _{R}(b)+i\chi _{I}(b)~.
\end{equation}%
Separating real and imaginary parts 
\begin{equation}
1-\cos \chi _{R}\ e^{-\chi _{I}} = \frac{1}{\sqrt{\pi }}\widetilde{T}_{I}(b)  \label{cosR} \\
\end{equation}
and 
\begin{equation}
\sin \chi _{R}\ e^{-\chi _{I}}  = \frac{1}{\sqrt{\pi }}\widetilde{T}_{R}(b)  \label{sinR}
\end{equation}%
we obtain 
\begin{equation}
 \chi_I(b)= -\frac{1}{2}\log\bigg[\frac{1}{\pi}\bigg(\widetilde{T}_{R}(b)^2+(\widetilde{T}_{I}(b)-\sqrt{\pi})^2\bigg)\bigg] ~.  
\label{chi_I}
\end{equation} 
so that the b-unitarity condition in Eq.(\ref{b-unitarity}) reads simply 
\begin{equation}
\chi_I(s,b) \geq 0   ~ ~~   , ~      \forall s,b  .  
\label{chiI}
\end{equation}

With monotonic behavior of the scattering amplitudes, our solutions
are restricted to the branch where $\chi _{R}\geq 0$.  
We need special care to write the expression for $\chi_R$, because it enters the second quadrant for small $b$. 
At the point $b=b_{\rm root}$ where $\widetilde{T}_{I}(b_{\rm root} )=\sqrt{\pi}$,  $\cos{\chi_R}$ becomes zero, 
and it is negative between $b=0$ and $b=b_{\rm root}$.  To have continuity, avoiding  that a calculator produces 
a positive value in the fourth quadrant, we must write the function $\arctan$ with two arguments. In the  
form used by the Wolfram Mathematica software, we write 
 \begin{eqnarray}
\chi_R(b)&=& \arctan[ (\sqrt{\pi}-\widetilde{T}_{I}(b)),\widetilde{T}_{R}(b)]   \\
  & = &  \frac{\pi}{2}-\arctan[\widetilde{T}_{R}(b),\sqrt{\pi}-\widetilde{T}_{I}(b)] ~. \nonumber 
\label{chiR}
\end{eqnarray}
 
\begin{figure*}[b]               
                        \includegraphics[width=8cm]{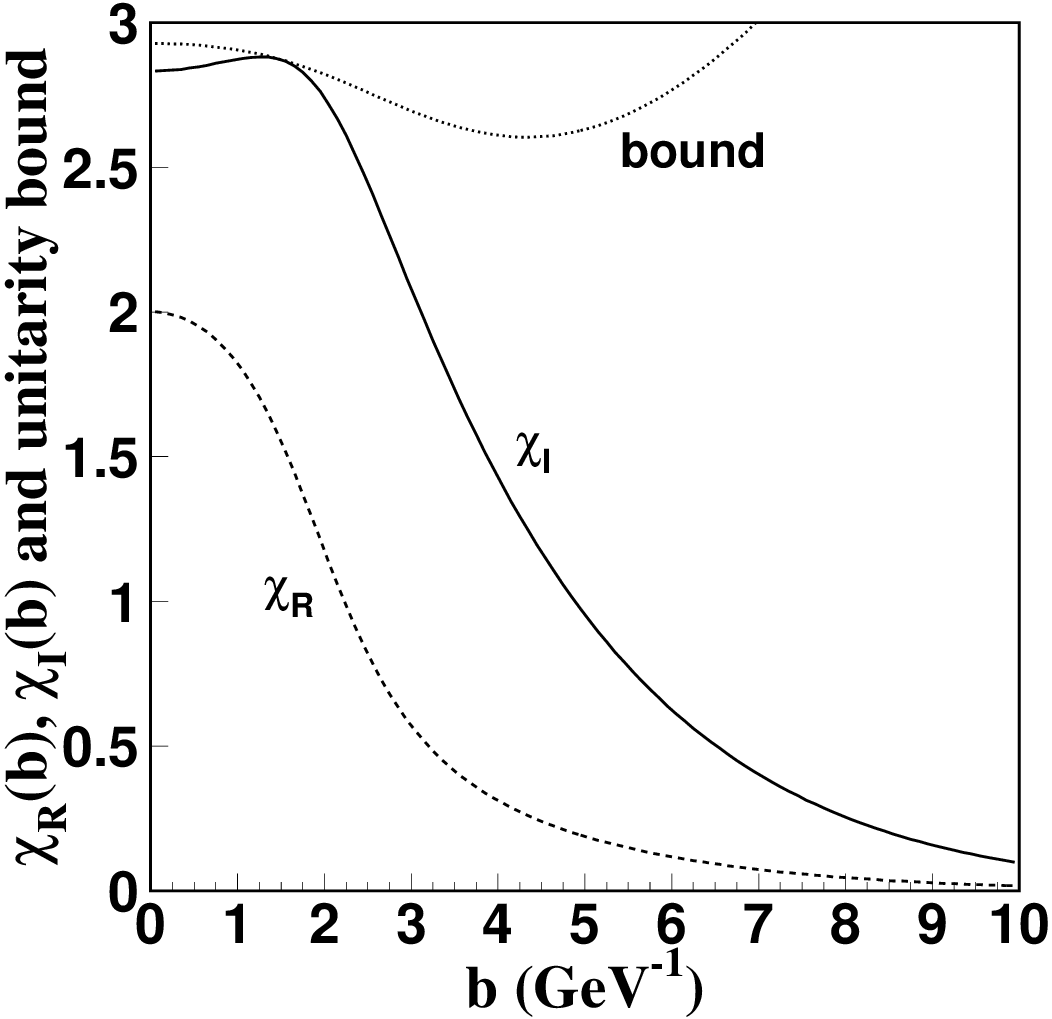} 
                    \includegraphics[width=8cm]{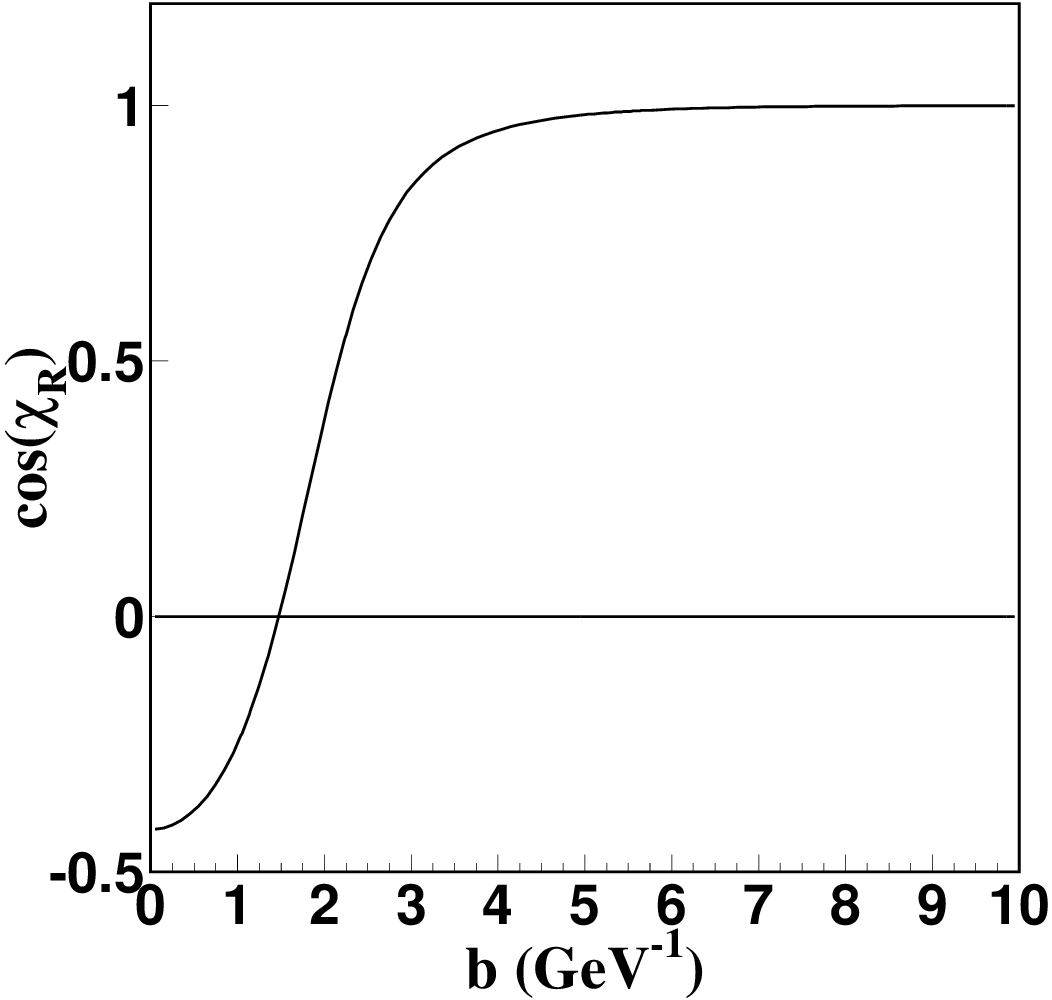} 
        \caption { Eikonal quantities.
      a) The quantity  ${\it bound}(b) = -(1/2) \log(T_R^2/\pi) $ 
         shown with dotted line  participates in the  constraint  of Eq.(\ref{bound});      
          b) $\chi_R(b)$ is in the second quadrant for small $b$ , with  $\chi_R(0)=2.0010 $ ,
      and $\cos{\chi_R(0)}=-0.4170$  ; at $b=b_{\rm root}=1.4739~\GeV^{-1}$, we have 
 $\chi_R(b_{\rm root})=\pi/2$ and  $\cos{\chi_R} (b_{\rm root})=0$  ~. 
           \label{eikonals}  }  
         \end{figure*}

In terms of the eikonal function, we have
\begin{eqnarray}
\frac{d\widetilde{\sigma }_{\mathrm{el}}(s,\vec{b})}{d^{2}\vec{b}}
&=&1-2\cos \chi _{R}~ e^{-\chi _{I}}+e^{-2\chi _{I}}, \\
\frac{d\widetilde{\sigma }(s,\vec{b})}{d^{2}\vec{b}} &=&2\left( 1-\cos \chi
_{R} ~ e^{-\chi _{I}}\right)   ~ ,  \\
\frac{d\widetilde{\sigma }_{\mathrm{inel}}(s,\vec{b})}{d^{2}\vec{b}}
&=&1-e^{-2\chi _{I}} ~ .
\label{b-diff}
\end{eqnarray} 
These expressions are plotted in Fig.\ref{b-amplitudes}, and the explicit 
representation for   $\cos{\chi_R}$  and the  expression   in Eq.(\ref{chi_I}) 
for $\chi_I(b)$ are  plotted in Fig.\ref{eikonals}.  

The function  $\chi_I(b)$  is not monotonically decreasing, 
starting with
$\chi_I(0)=2.83208$ , and presenting a maximum at $b_{\rm max}=1.2700 ~\GeV^{-1}$
with value $\chi(b_{\rm max})=2.8818$. 
This  property is not observed in our previous analyses at lower energies $\sqrt{s}\leq 7$  TeV, 
where $\chi_I$  is always monotonically decreasing function in $b$. 
To detail this peculiar behaviour,
and illustrate the effect of the real part, 
we point out that from Eqs. (\ref{cosR},\ref{sinR}) together with Eq.(\ref{chiI}) we have
\begin{equation}
0 \leq \chi _{I}(b) \leq  -\frac{1}{2}\log \big( \frac{\widetilde{T}_{R}(b)^{2}}{\pi} \big)~ \equiv {\it bound}(b)~.
\label{bound}
\end{equation}%
The expression  ${\it bound}(b) $ 
is plotted in dotted line in Fig.\ref{eikonals}a. At $b=b_{\rm root}=1.47393 \GeV^{-1} $,  where 
$ \widetilde{T}_{I}(b)-\sqrt{\pi} = 0$ , $\chi _{I}(b)$ touches ${\it bound} (b)$. Everywhere
else, the inequality holds. This happens even at the maximum 
$b_{\rm max}$ of  $\chi_I(b)$, where ${\it bound}(b_{\rm max})=2.8895$ is slightly  larger
than $\chi_I(b_{\rm max})$=2.8818.


It is interesting   that the differential inelastic cross section
$ d\tilde{\sigma}_{\rm inel}/d^{2}\vec{b} $
in Fig.8-c is almost fully saturated
($\simeq1$) in the central collision region up to $b<4\GeV^{-1} 
\approx 0.8$ fm. This can be seen also from the behavior of $\chi
_{I}(b)$ in Fig.9, where for $b<4$, it is  $\chi_{_{I}}>1.5$ so that $\exp\left(
-2\chi_{_{I}}\right)  \leq0.05$. In the classical picture, from the central to
approximately the half overlap impact parameter, the pp system behaves as 
completely absorptive, leading to  particle production channels. We note,
however, that this does not mean that the elastic differential cross section
$d\tilde{\sigma}_{\rm el}/d^{2}\vec{b}$ is null, due to the wave nature of the
scattering. The diffractive wave as the reflection of inelastic scattering
contributes to the elastic channel with almost the same magnitude as the inelastic
one inelastic one even for the extreme case of a black disk.
  
On the other hand, at this energy, we note that the 
elastic scattering profile at $b=0$ 
is rather large, exceeding the inelastic profile, which was never observed in our previous 
analyses. Furthermore, we also observe for the first time, a small decrease the inelastic 
profile near $b=0$ (almost invisible in Fig.  8, as direct reflection of the behavior of 
$\chi_I$ shown in Fig.  9). We will return to this point later.

As claimed in previous studies, in the very peripheral collisions
(at this energy, $b \geq 8 \GeV^{-1} \simeq 1.6 ~{\rm fm}$), 
   contributions from elastic processes become negligible and 
inelastic processes are dominant.
 
Physically  speaking, this part can be associated to diffractive particle production 
mechanism. In  $b$-space, this constitutes a rather diffused surface structure 
with a long tail 
 in $d\tilde{\sigma}_{\rm inel}/d^{2}\vec{b}$.  We may associate such processes
(forward scattering) with those from the excitation of the vacuum through the
non-perturbative processes. In \cite{KFK_3} we argue that the existence of such a
long tail in $d\tilde{\sigma}_{\rm inel}/d^{2}\vec{b}$ and vanishingly small
values of $d\tilde{\sigma}_{\rm el}/d^{2}\vec{b}$ for large $b$, 
say $>9  ~ \GeV^{-1}$, can
be considered as responsible for the ratio, $\tilde{\sigma
}_{\rm inel}/\tilde \sigma$ being significantly larger than that of a black-disk limit,
namely $1/2$.  There, assuming the geometric scaling property for $d\tilde{\sigma}%
_{\rm inel}/d^{2}\vec{b}$, we extrapolated this ratio to 13 TeV, predicting the
value
\[
\left(  \tilde{\sigma}_{\rm inel}/\tilde \sigma\right)  _{\rm extrapolation}=0.7428,
\] while the present analysis gives
\[
\left(  \tilde{\sigma}_{\rm inel}/\tilde \sigma\right)_{\rm Totem}=0.7212,
\]
which is $3\%$ smaller, but yet definitely far from the black disk limit.    

 

 \section{Energy Dependence \label{energy}}  


The KFK model represented by Eqs.(\ref{b-AmplitudeN},\ref{Shape-b}), or alternatively 
Eq. (\ref{hadronic_complete}) in $|t|$ space, has been used in the description of $d\sigma/dt$ 
data    at several energies,   
and its properties and predictions in both  $|t|$ and $b$ spaces were studied also for 
 cosmic ray showers \cite{CR_2014}. 
 
Data of pp elastic scattering  covering regularly  from small  to large $|t|$  
 are available  in the ISR range (up to 63 GeV), and at 7 and 13 TeV in  LHC Totem measurements. 
The  comprehensive analysis  of all $d\sigma/dt$  data then available (up to $\sqrt{s}$=7 TeV) was 
made \cite{KFK_3} with a study  of the energy dependence of the KFK parameters, including 
predictions for  13 and 14 TeV.

The  13 TeV data are  more precise and cover wider $|t|$  range than the 7 TeV data, 
allowing realistic determination of the   amplitudes in KFK model.
This is the purpose and the achievement of the present work. 
The results obtained  lead to revision and extension of the previous 
analysis, and the updated revision is presented in this section.

We  stress that KFK provides a framework that is particularly important for a study of 
the real part, that is elusive in the forward region, becoming  influent at mid  $|t|$ 
and dominant after about  3 $\GeV^2$.  Due to the small value of $\rho$,   the interplay 
of the  electromagnetic and  real part of the nuclear amplitude is very delicate.
 A detailed analysis of the forward data at 8 TeV \cite{TOTEM_8_TeV},  accounting for 
the role of the real part of the hadronic amplitude in the CNI  contribution,  has 
demonstrated  the 
importance of the hadronic model in the determination of the forward scattering parameters, 
leading to the values $\rho= 0.12\pm0.03$ and $\sigma=102.9\pm2.3$. Particularly the 
$\rho$ value, small compared to 0.14 of  COMPETE preference, anticipated the tendency
that was later confirmed in measurements at 13 TeV. The $\rho$ value at 8 
TeV   affects  the revision of  parameters  presented  in this section,
particularly leading to $\rho=0.115 \pm 0.001$  at 7 TeV, with very good $\chi^2$.
The   decisive influence of the hadronic amplitudes in the study of the 
phase in the Coulomb-Nuclear Interference, with consequences in the evaluation of $\rho$ 
and $\sigma$, was also demonstrated at 13 TeV \cite{PLB2019}.  

 Table \ref{tablefive}   shows the optimal values of the model parameters  for $\sqrt{s}=52.806 $ GeV, chosen   as
representative of the  ISR range,  together with the    results of 7 TeV and 13 TeV.  
  The table   introduces  an alternative notation, 
defining  quantities  $\eta_K$  through 
\begin{equation}
\eta_K= \gamma_K a^2  
\end{equation} 
that have the same $\GeV^{-2}$ units as the other six quantities $\alpha_K$, $\beta_K$, $\lambda_K$, 
used instead of the dimensionless $\gamma_K$  used in the text and in previous work.  
  {\small     
\begin{table*}[ptb]
\caption{ Parameters of the amplitudes in the KFK model determined at the energies 52.806 GeV, 7 TeV and    13 TeV. 
 For uniformity, the  table   uses the alternative parameters $\eta_K= a^2 \gamma_K $ with units $\GeV^{-2} $ instead of 
the dimensionless $\gamma_K$.  Notice   values of  $a^2$ for different energies.
 For 0.0528 TeV the data reaches $|t|= 10 \GeV^2$ and the three-gluon egxchange term is included. }
   \label{tablefive}
   \begin{tabular*}{\textwidth}{@{\extracolsep{\fill}}cccccc|cccc|cccc@{}}
     \hline
 \hline
$\sqrt{s}$&$a^2$&N&$\chi^2$& $\sigma$ &$\rho$ &$\alpha_I$ &$\beta_I$  &$\lambda_I$&$\eta_I $  &$\alpha_R$ &$\beta_R$  &$\lambda_R$&$\eta_R $  \\ 
 TeV   & $\GeV^{-2}$&  &    & mb &     & $\GeV^{-2}$&$\GeV^{-2}$&$\GeV^{-2}$&$\GeV^{-2}$&$\GeV^{-2}$&$\GeV^{-2}$&$\GeV^{-2}$&$\GeV^{-2}$ \\ \hline
0.0528&1.39  & 97 &0.9251 &42.54 & 0.078 ~   & 5.958     &2.348      & 9.451     &10.5778    & 0.0710    & 1.144     &  1.131    & 11.794    \\ \hline 
 7    & 2.00 &165 &0.2957 &98.75 & 0.115 ~& 13.730    &4.100      &22.040     &16.3000    & 0.2572    & 1.405     &  3.856    & 15.576   \\ \hline 
13    &2.1468&428 &1.567  &111.56 &0.118 ~ & 15.701    &4.323      &24.709     &16.7858    & 0.2922    & 1.540     &  4.472    & 16.107   \\ \hline 
\end{tabular*}
\end{table*}  

Using the forms $\eta_K$ instead of  $\gamma_K$, the non-perturbative shape functions are written
   \begin{equation}
\widetilde{\psi }_{K}(s,b)=\frac {  2  e^ {\big(\eta_{K} - \sqrt{\eta_{K}^{2}+{b^{2}}{a^2}}\big)/a^2 } }
{    \sqrt{\eta_{K}^{2}+{b^{2}}{a^2}}    }
   \Big[1-  e^ {\big(\eta_{K} -  \sqrt{\eta_{K}^{2}+{b^{2}}{a^2}}\big)/a^2 } 
                                          \Big]~.  \label{Shape-b_2}
 \end{equation}
Consequently,  in  $t$-space the shape function obtained by Fourier Transform is written
\begin{eqnarray}
 \label{psi_st_2}  
&&\psi _{K}(\gamma _{K}(s),t) \\
&=&2~e^{\eta _{K}/a^2}~\bigg[
{\frac{ {e}^{-(\eta _{K}/a^2)\sqrt{%
1+a^2|t|} }}{\sqrt{1+a^2 |t|}}}-  \mathrm{e}^{\eta _{K}/a^2}~{\frac{e^{-(\eta
_{K}/a^2)\sqrt{4+a^2|t|}}}{\sqrt{4+a^2 |t|}}}               \bigg]~.    \nonumber
\label{t-Ampl-2}
\end{eqnarray} 
 Using these sets of  values, we updated the energy dependence 
of the KFK parameter values   \cite{KFK_3}  as
 \begin{eqnarray} 
&&  \alpha_I= 8.97889 + 1.87838\log{\sqrt{s}}+ 0.289432\log^2{\sqrt{s}} \\
&&  \beta_I=3.40059+0.35881\log{\sqrt{s}}+0.000315\log^2{\sqrt{s}} \\     
&&  \lambda_I = 15.22340  + 2.88969\log{\sqrt{s}} + 0.3152\log^2{\sqrt{s}} \\
&& \alpha_R  = 0.16377 +0.04144\log{\sqrt{s}}+ 0.003365\log^2{\sqrt{s}} \\  
&& \beta_R=1.13041+0.083146\log{\sqrt{s}}+0.029841\log^2{\sqrt{s}} \\
&& \lambda_R  = 2.31722+0.63662\log{\sqrt{s}}+0.079324 \log^2{\sqrt{s}}\\   
&& \eta_I= 13.79950 + 1.13219 \log{\sqrt{s}}+0.012510 \log^2{\sqrt{s}}        \\   
&& \eta_R =13.98229+0.78910\log{\sqrt{s}}+0.015369\log{\sqrt{s}}      
\label{coefficients} 
\end{eqnarray} 
with $\sqrt{s}$ in TeV  and units $\GeV^{-2}$ for all quantities.  
These forms as functions of the energy  are  shown in Fig.\ref{parameters-fig}.
  
In the table we notice that the correlation length squared $a^2$, serving as scale 
for the gluon correlations in the transverse collision plane for the nonperturbative term,  
has regular energy dependence, staying close to the   
    value obtained in static lattice calculation. 
The   KFK amplitudes for   $d\sigma/dt$ are  sensitive to these  values, and  
a representation appropriate   for interpolation  is 
\begin{equation} 
a^2 = 1.64036 + 0.145122 \log{\sqrt{s}}+0.0204  \log^2{\sqrt{s} }   ~   ~ \GeV^{-2}   .       
\end{equation}

\begin{figure}[b]
  \includegraphics[width=8.0cm]{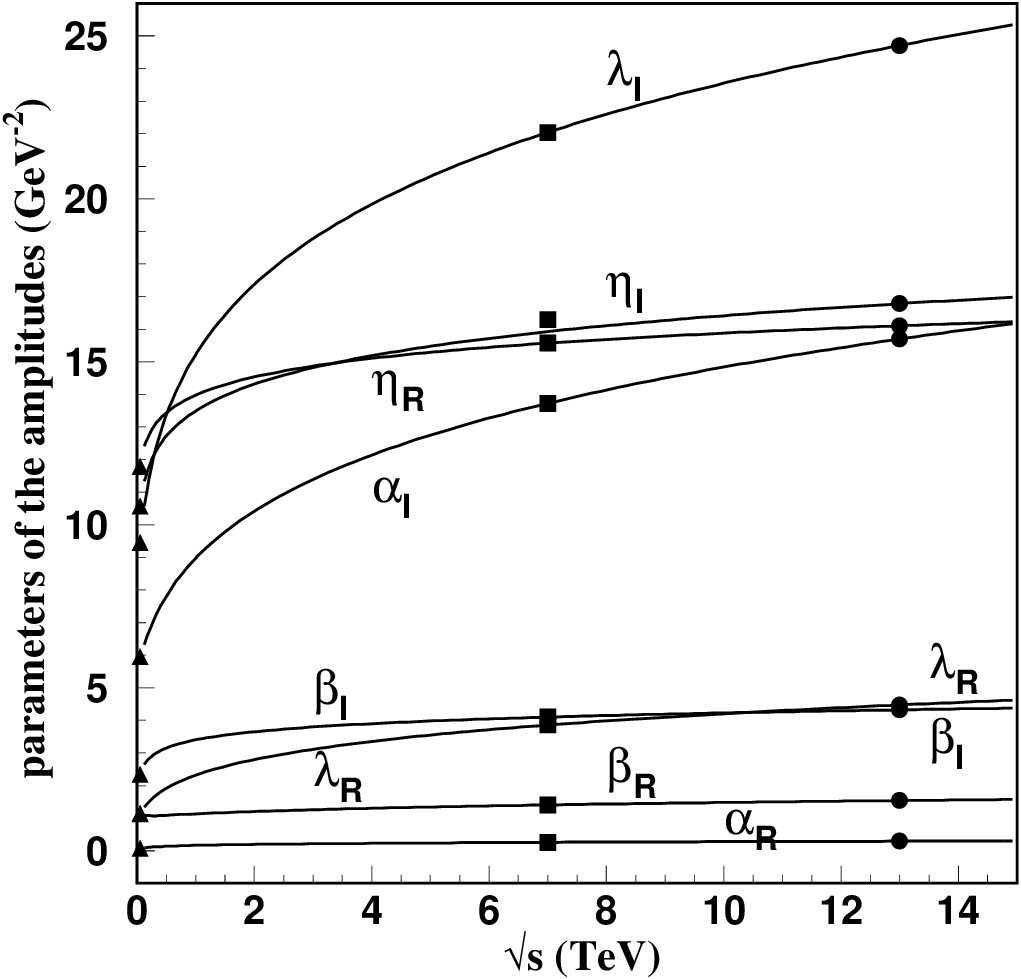} 
 \caption{  \label{parameters-fig}    Energy dependence of the KFK parameters,    
obtained by direct analysis of  $d\sigma/dt$  data at ISR energies, and at 
LHC energies 7 and 13 TeV. 
  The dots mark  the values of the parameters  at
 reference values 52.8 GeV, 7 TeV and 13 TeV, as given in Table \ref{tablefive}. 
}
 \end{figure}  

The total cross section reads 
\begin{eqnarray} \label{sigtotal}
  \sigma & = &2.7606 (\alpha_I+\lambda_I)    \\ 
& = & ~ 66.8129+13.1627\log{\sqrt{s}}+1.6691\log^2{\sqrt{s}} ~ ~~  {\rm {mb}} ~,  \nonumber 
 \end{eqnarray}  
and we recall that $\rho=(\alpha_R+\lambda_R)/(\alpha_I+\lambda_I) $ is given 
by Eq.(\ref{rho_par}).  

The  slopes of the amplitudes, shown in Fig.\ref{slopes-fig}, can be represented by simple forms 
\begin{eqnarray} \label{slopes}
  B_I&=& 17.270+1.457 \log{\sqrt{s}}+0.006 \log^2{\sqrt{s}}     \\
  B_R&=&22.457+1.356\log{\sqrt{s}}+ 0.070 \log^2{\sqrt{s}}   \nonumber  .
\end{eqnarray} 
with units  $\GeV^{-2}$.
The structure of the forward amplitude with  different slopes $B_I$ and $B_R$ is 
crucial in the analysis of the CNI range for determination of $\sigma$ and $\rho$.   
The stronger real slope $B_R$ indicates the presence of the close zero predicted by 
Martin's theorem.
\begin{figure}[b]
  \includegraphics[width=8.0cm]{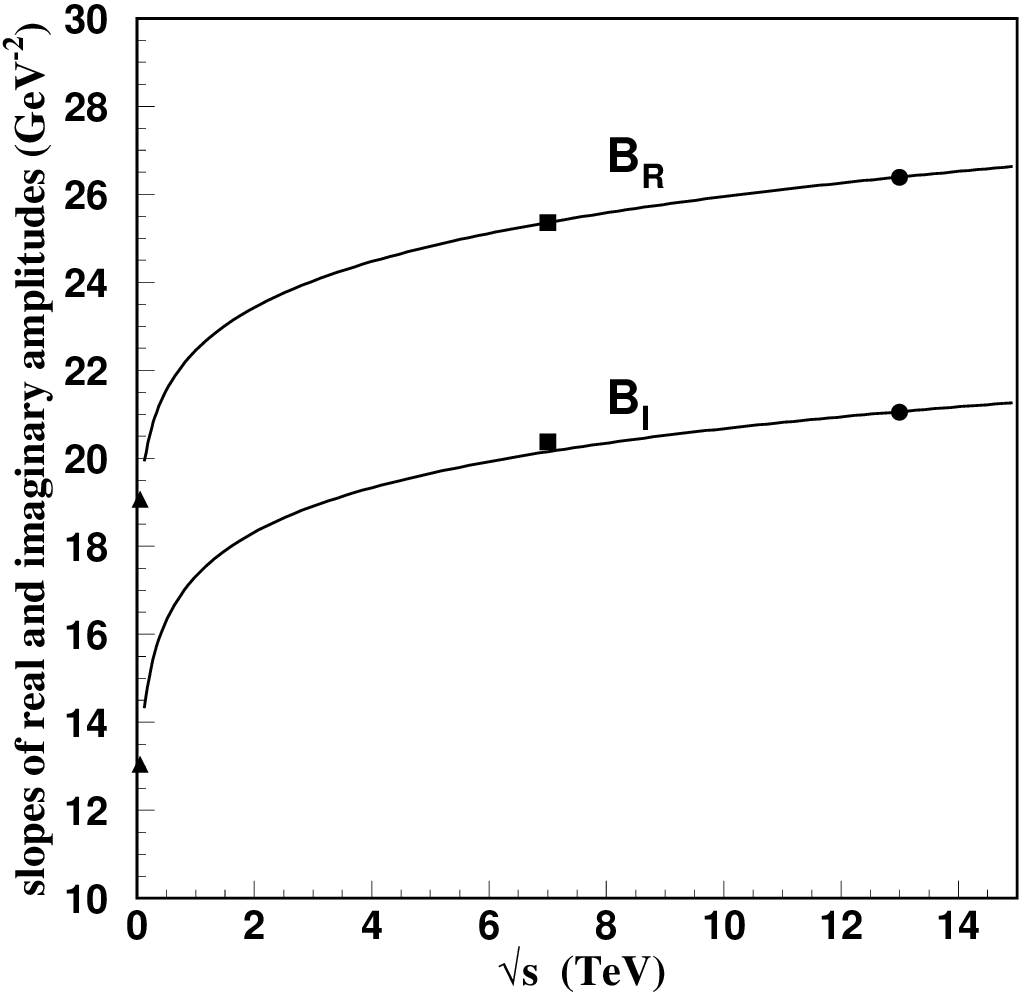} 
 \caption{ Energy dependence of the slopes of the real and imaginary amplitudes. 
  \label{slopes-fig} }  
\end{figure}   
  
It is interesting to observe the energy dependence of properties of the amplitudes in $b$-space
\cite{KFK_3,CR_2014}. Fig.\ref{eikonals} shows that at 13 TeV the elastic  differential cross 
section at $b=0$ is larger than the inelastic
quantity, while at lower energies  the inverse is true.  According to our description, 
the ratio elastic/inelastic  at $b=0$ increases with the energy, with values 0.56, 0.90, 
1.0, 1.06  for 52.8 GeV, 7 TeV, 10.57 TeV   and 13 TeV  respectively.  The energy 
dependences are determined with 
$d\sigma/dt$  data that have a wide  coverage 
in t and permit to
obtain the parameter values with excellent precision for each given energy 
up to 13 TeV. However, the forms 
have limited local validity, like  Taylor expansions 
in $\log{\sqrt{s}}$ up to second order, and are not  
  adequate for  extrapolation to very high energies.
Nevertheless, it is tempting to compare the predictions resulting from
the present analysis to, for example, a cosmic ray energy scale, as $\sqrt{s}$ = 50 TeV. 

The above mentioned ratio elastic/inelastic at $b=0$ increases as high as 1.56, while 
Eq.(\ref{sigtotal})   predicts  $\sigma= 143.85 $ mb  at  50 TeV,
that is consistent with the estimated values of sigma(pA) data \cite{CR_2014}. 
Values of some derived quantities are shown in Table \ref{tablesix}. 
  
Fig.\ref{cross-b-50} shows the elastic, inelastic and total differential cross 
sections in  $b$-space  for 13 and 50 TeV. 
In view of the study of properties of the terms of the amplitudes in subsection \ref{b_amplitudes}  
we learn that this increase of the elastic cross section at $b=0$ is mainly due to the perturbative
terms. However, we must remark that the range around $b=0$ is reduced in the  $ bdb $ integration,
and that the inelastic cross section dominates for larger $b$, so that the integrated inelastic is larger 
than the integrated elastic at all energies. 
The ratios are given in Table \ref{tablesix}.

\begin{figure*}[b]
    \includegraphics[width=8cm]{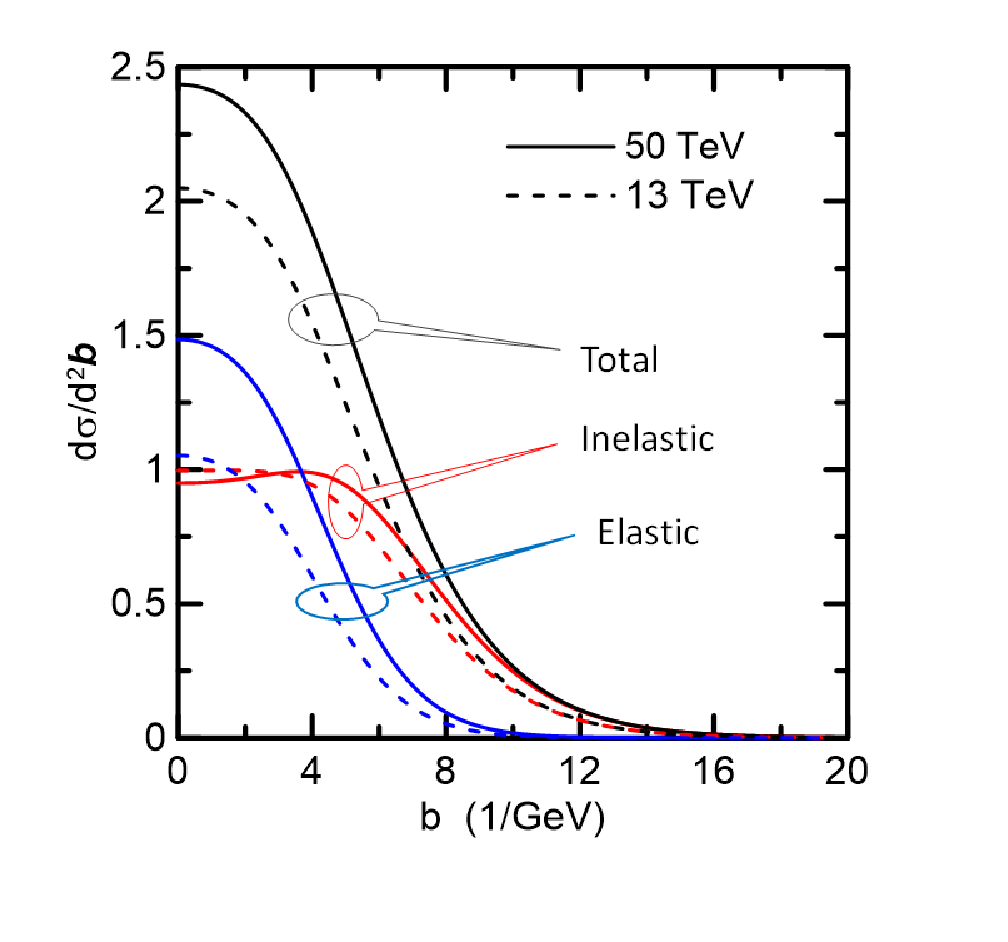}   
    \includegraphics[width=8cm]{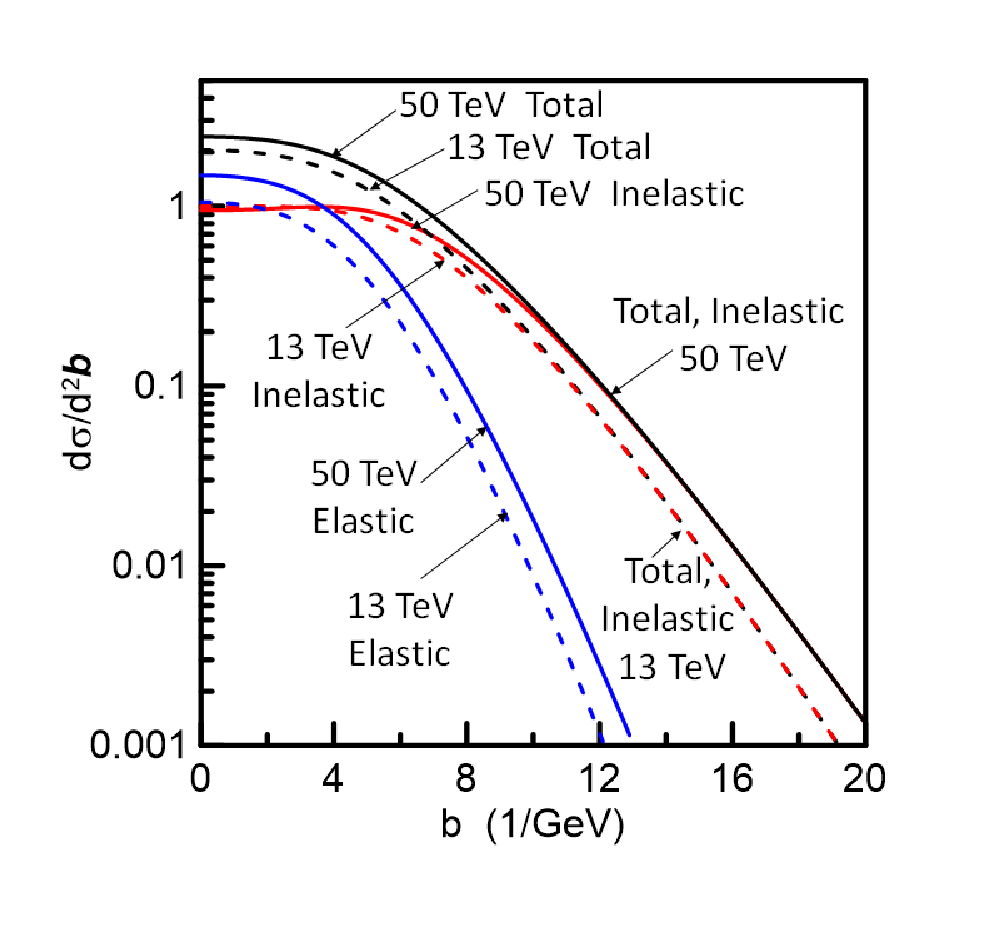} 
  \caption{Differential cross sections in $b$-space at energies 13 and 50 TeV.
At $b=0 $ the ratio of differential cross sections elastic/inelastic 
increases  from 1.06 at 13 TeV  to 1.56 at 50 TeV. As the energy increases, the 
interaction at the center of the proton becomes increasingly elastic.
On the other hand, the integrated cross section is dominantly inelastic,
as its range is more extended and the value is favored by the $b$  factor in the 
integration.  For more clarity, the figure is repeated in log scale. 
  \label{cross-b-50} }
\end{figure*}

In Fig.\ref{cross-b-50} we observe that the inelastic differential cross section is never saturated (namely it is always smaller than 1, with the eikonal $\chi_I$ larger than zero), 
while the elastic and 
total  quantities are strongly enhanced in the region close to $b=0$.   
For very central collisions, with the impact parameter smaller than
the nucleon geometric size, inelastic processes at 50 TeV  are visibly 
 suppressed compared to the 13 TeV case. 
To be more precise, such suppressions of inelastic profile near $b=0$ already 
started in the 13 TeV data, although not being quite visible in the figure.
 However from the behavior of $\chi_I$ in Fig.\ref{eikonals} near $b=0$, 
together with Eq.(\ref{b-diff}), it is clear that the inelastic profile has a minimum at $b=0$.   
as mentioned in the previous section. 
The ratio   elastic/inelastic cross sections at $b=0$ increases fast   because the 
elastic part increases and simultaneously the inelastic part decreases. 
For much higher energies, this tendency is more enhanced.

The concept of the impact parameter  $b$ is classical and cannot 
be associated  with a real physical  observable in microscopic systems.
Nevertheless, the present results suggest an image that, at ultra high energies, 
the two colliding protons tend to behave as two thin, inter-penetrable hard disks 
so that the process becomes elastic scattering dominant, decreasing the inelastic channel. 
This seems to occur in the $b$-domain  corresponding to the proton radius 
($b<{\rm R_{proton}} \approx 0.85 ~ {\rm fm}$). Such image may require the existence 
of some non-causal transverse correlation between the whole colliding protons, for 
example similar to the exclusion principle. 
It will be interesting to compare the elastic differential cross section for small 
$b$ in pp and $\rm{ p \bar p}$ collisions. If no such enhancement appears in 
$\rm{ p \bar p}$, a simple idea of exclusion principle may  be compatible, 
although sometimes  the differences of scattering amplitudes in pp and   
$\rm{ p \bar p}$ are considered as signal of odderon existence. 
For larger $b$ our model indicates that the cloud of vacuum fluctuations 
around the proton dominates the process,  contributing to the inelastic 
(particle production) channels.  
 
 These considerations  show that precise data on the scattering amplitude for different 
values of $\sqrt{s}$ and with  wide $|t|$ range are necessary for the  understanding of
the structure of proton and of   the surrounding QCD field  in the 
collision region.

 {\small
\begin{table*}[ptb]
\caption{Quantities derived from the energy dependence expressed by the 
interpolation equations (\ref{coefficients}).   
The quantities $Z_I$,  $Z_R^{(1)}$  and $Z_R^{(2)}$ are the locations ($|t|$ values) of the zeros 
of the imaginary and real amplitudes, that are important in the dip-bump structure. The 
integrated cross sections do not show tendency for  black disk collision.     
   \label{tablesix}     }
    \begin{tabular}{@{\extracolsep{\fill}}ccccccccccccc@{}   }
      \hline
\hline
 $\sqrt{s}$ &$Z_I$ &$Z_R^{(1)}$ & $Z_R^{(2)}$&$\rho$& $\sigma_{\rm tot}$&$\sigma_{\rm el}$&$\sigma_{\rm inel}$ &$\sigma_{\rm el}/\sigma_{\rm inel}$&$\sigma_{\rm el}/\sigma_{\rm tot}$ & $|t|_{\rm dip}$ & ${\rm h_{dip}}$     \\ 
     TeV    & $\GeV^{2}$  &$\GeV^{2}$  &$\GeV^{2}$   & $          $    & mb  & mb  & mb  & &   &$\GeV^{2}$  & mb/$\GeV^2$      \\ \hline
  $7$      &$0.479$&$0.209$&$1.144$&$0.115$ & $98.75$ & $25.37 $  & $ 73.38$ &$  0.346$& 0.257 & 0.487 & 0.012   \\ \hline 
  $13$     &$0.460$&$0.200$&$1.180$&$0.118$& $111.56$ & $31.10$ & $80.46$ &$ 0.386 $& 0.279 & 0.470    & 0.026    \\ \hline 
  $20$     &$0.453$&$0.195$&$1.218$&$0.120$& $121.22$ & $35.56$ & $85.66$ &$ 0.415$&0.293 &  0.460     & 0.032   \\ \hline 
  $50$     &$0.428$&$0.183$&$1.345$&$0.123$ & $143.85 $& $47.59$ & $96.26$ &$0.494 $& 0.331 & 0.442    & 0.051   \\ \hline 
\end{tabular}
\end{table*}      }   


\section {  Other Models  \label{MODELS}      }

The present paper is mainly dedicated to the analysis of the $|t|$ dependence 
of pp elastic scattering measured at 13 TeV, characterized by  
unique statistical  quality and wide   $|t|$ coverage.
These data brought surprises and oportunities 
for   theoretical models. Several well established frameworks revised their 
assumptions and results. The response of the proton in the scattering process
may change because Lorentz contraction  puts the partons closer,
 and correlations (and even exclusion principle) act differently
as  energy increases. 

  In the present work  KFK model  gives high precision representation for 
all data with identification of the   real and imaginary  amplitudes, 
and shows $\chi^2$ values for separate  ranges with a unique solution,
both with statistical and with  combined statistical and systematic errors. 
  Although consistent and detailed, the significance of the results depends 
on the analytical forms used,  and it is important to compare our calculations 
with the results obtained in  different frameworks, trying to learn about the 
  meaning of each one. 

 Comprehensive and competent reviews are available, discussing  several  aspects 
of pp elastic scattering,
 in both $s$ and $t$ variables  \cite{Fiore,Pancheri}. In this section we mention
some specific calculations that deal with  aspects related with the  present work.

\subsection{Pomeron Models \label{pomeron_section} }


Models based on Regge formalism are traditional in studies of hadronic scattering, 
giving connection between the $s$ and $t$ variables  in  forward scattering 
for many hadronic systems in terms of kinematical forms called Regge trajectories.

To describe the observed curvature in the diffractive peak of pp scattering, the 
main Pomeron trajectories must become non-linear, and modulated forms  with adjustable 
parameters are proposed. To extend the use of  Regge models up to the 
dip, the hadronic amplitude must have a zero, and terms of negative sign must be 
included in the framework. Thus the contribution of the exchange of two Pomerons 
\cite{DL2,DL3} is introduced, with formalism and  parameters adjusted to locate 
the dips and estimate their heights. We are not aware  that this has been 
achieved with good accuracy, but the conclusion of these two papers is that at 13 TeV there 
is not evidence for   an Odderon contribution in this framework.  In an alternative
approach
\cite{Szanyi}, without two-pomeron exchanges, 
Pomeron and Odderon terms are added on equal foot, both with 
double poles and independent parameters. The very forward CNI range is not  
treated, but the description of the  dip/bump region at 13 TeV 
is satisfactory ($\chi^2$ value for this specific range is not informed), up to 
$|t|\approx 2.0 ~\GeV^2$. 

We emphasize that in this Regge framework  
the  data for large $|t|$ (say $|t|\geq 2.5 \GeV^2$) range  
   are  not properly represented.  This is evidence of the absence of  
 knowledge of the  transition from soft to hard dynamics,   
possibly with perturbative three-gluon exchange influencing the tail region
and shows the need  for more measurements.

Corresponding to these two  approaches, namely   two-pomeron exchange 
(also multi-pomeron exchanges) and added odderon exchange, the additional  
terms  with negative sign leading to dip and bump, are accounted for 
equivalently in the non-perturbative shape functions 
of KFK, that guarantee  these properties of the amplitudes.

A more recent work \cite{Godizov}  explores the Regge framework, introducing the 
tradicional soft Pomeron with nonlinear trajectory  and the hard Pomeron with 
stronger slope. These quantities are added in an eikonal approximation. The parameters 
adjusted to include the 13 TeV data allow a good representation of the pp data for 
7 TeV and 13 TeV, particularly for large  $|t|$, and the authors inform  that the 
hard Pomeron pole is crucial in this aspect. No Odderon presence is claimed here. 
The  $|t|$ space amplitudes of  in this calculation are similar the KFK amplitudes.

 Broilo, Luna and Menon    \cite{Luna} studied 
 the energy dependence of $\sigma(s)$ and $\rho(s)$ including  the 
13 TeV data in the statistical analysis of all data from $\sqrt{s}=5~ \GeV$  reported 
by the Particle Data Group (PDG), investigating comparatively the contributions of 
powers and/or logarithms in the Pomeron exchange terms \cite{Luna}.   
The conclusion favors the  choice of the parametrization with $\log{s}$ and 
$\log^2{s}$ in $\sigma(s)$, excluding  power forms. At 13 TeV the parametrization leads to  
$\sigma(s)$ = 107.2 mb, that disagrees with the calculations  based on  $d\sigma/dt$ , 
   whereas leads to   $\rho=0.1185$ that agrees 
with   KFK value  for zero Coulomb interference phase. 

Unfortunately, the  determination  of the $|t|=0$ quantities such as 
 $\sigma(s)$ and $\rho(s)$ 
based purely on the bare data   of PDG  is not  secure, because  this 
inclusive data basis 
  has not been not submitted to a selection  and evaluation of 
consistency and quality \cite{LOW_ENERGIES}. 
  Values of  $\sigma$ and $\rho$ are not  quantities directly   measured,  
 but rather are model dependent calculations, requiring identification of 
the imaginary and real parts of the amplitude, and   in many cases 
the $d\sigma/dt$ measurements  are not sufficient in range and quality for these 
calculations.   

\subsection{ Martin's Formula for the Real Part}

With basis on general principles of quantum field theory, A. Martin  obtained 
a formula \cite{Martin_2} connecting the real and imaginary parts of the complex 
amplitude of pp/${\rm{p \bar p}}$ elastic scattering. In principle the relation was 
established under restrictive conditions, as proximity of the asymptotic Froissart 
bound and limitation to the  very forward range.  The formula, that refers to the even 
component of crossing symmetry, includes also a scaling property incorporating 
 energy dependence in the relation. 
The  scaling property connecting $s$ and $t$ has been explored in several instances
\cite{JDD,Kohara,Pancheri_2}, describing  properties of the real and 
imaginary amplitudes in the forward range. 

Without considering Martin's formula as a theorem with strict constraints,  the 
relation was considered as a suggestion \cite{Menon_2}  for properties 
of the real part of the  full $|t|$ range data  of Fermilab and ISR 
experiments in the energy range $\sqrt{s}$ =19.4 - 62.5 GeV .  
The imaginary and real parts are fitted together, using a total 
of 12 parameters for each energy, with representations for real and imaginary parts
connected  by the formula. The numerical study includes also the 39 points of 
Faissler et al.  measurements \cite{Faissler} at 27.4 GeV, considered as universally valid  
for  the energy range investigated. The      
original Martin's real-part formula \cite{Martin_2} was used without the full scaling property,
namely it  is applied separately for each energy investigated, with determination of the best
parameters at each energy. 
The   fittings of the ISR data  show imaginary part with one zero and real parts with two 
zeros, just as we obtain in KFK model. 
 
The equation to be used is 
\begin{equation}
   T_R(s,t) = \frac{T_R(s,0)}{T_I(s,0)} ~  \frac{d}{dt} \big[t ~ T_I(s,t) \big] ~ . 
\label{Martin_t1}
\end{equation}
 Obviously ${T_R(s,0)}/{T_I(s,0)}=\rho$, but this quantity is not predicted by the formula, 
that specifically predicts the $|t|$ dependence of the ratio $T_R(s,t)/T_R(s,0)$ once 
the imaginary part $T_I(s,t)$ is given. 

To reproduce this study with the 13 TeV  data testing the KFK model, we do not fit freely the 
imaginary and real parts, but rather take $T_I(t)$ as known and obtain a prediction for
the real part by Martin's formula. We then write
 \begin{equation}
  \frac{T^{\rm Martin}_R(t)}{T^{\rm Martin}_R(0)}= \frac{d}{dt} \big[t ~ \frac{T_I(t)}{T_I(0)} \big] ~ .
\label{Martin_t}
\end{equation}
where  $T_I(t)$ is the KFK proposal treated  in Sec.\ref{data_analysis}.   
 In Fig.\ref{Martin_figure} we show KFK real amplitude normalized to one at the origin,
namely we plot $T_R(|t|)/T_R(0)$  from KFK (solid line) and 
${T^{\rm Martin}_R(t)}/{T^{\rm Martin}_R(0)}$ 
from Martin's formula
in Eq.(\ref{Martin_t}), with the given imaginary  ratio  $T_I(t)/T_I(0)$.  
 The important point 
for the KFK model is the confirmation of the properties of the amplitudes: one zero 
for $T_I(s,t)$  and two zeros for $T_R(s,t)$, with the real  part   
  dominant over the imaginary part after the bump. 
The  comparative plots   in  Fig.\ref{Martin_figure}  show that  
  differences  in    positions and   shapes.  
                 \begin{figure*}[b]               
              \includegraphics[width=8cm]{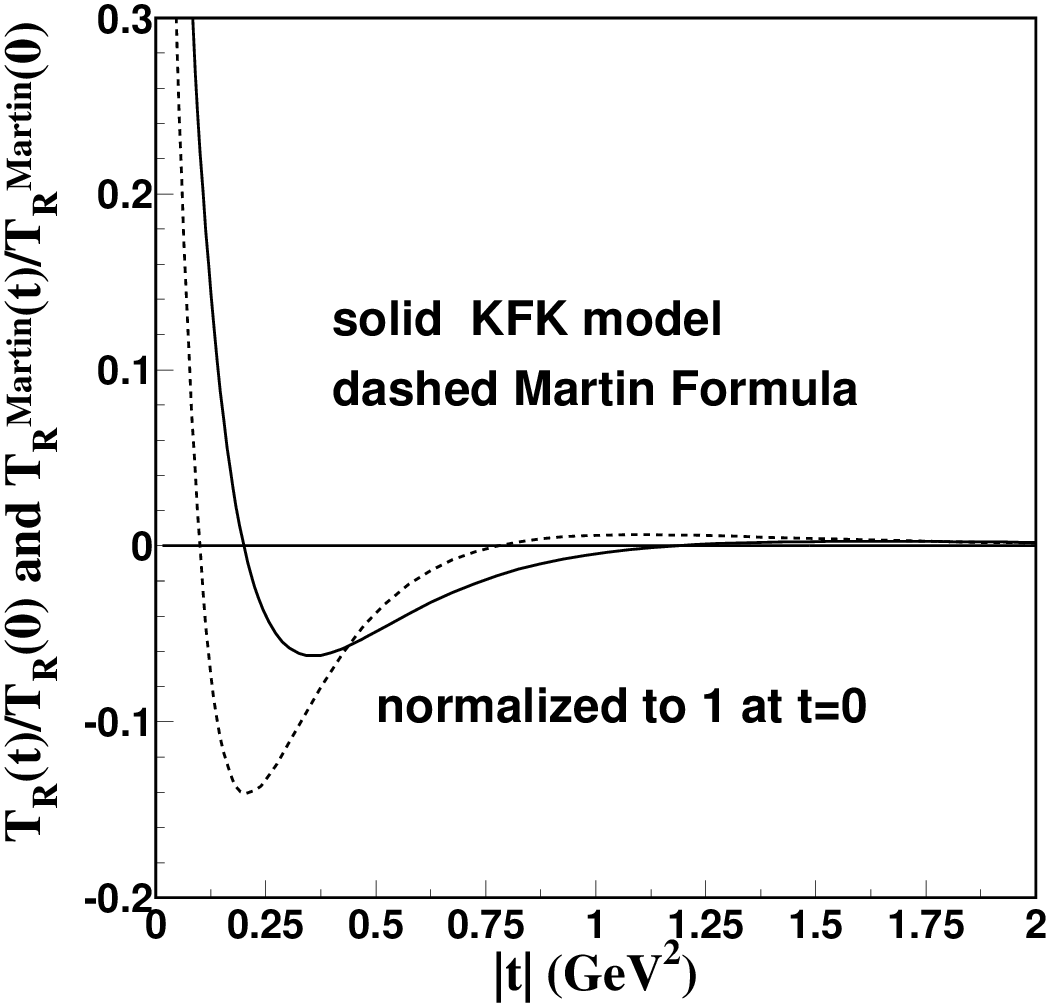} 
               \includegraphics[width=8cm]{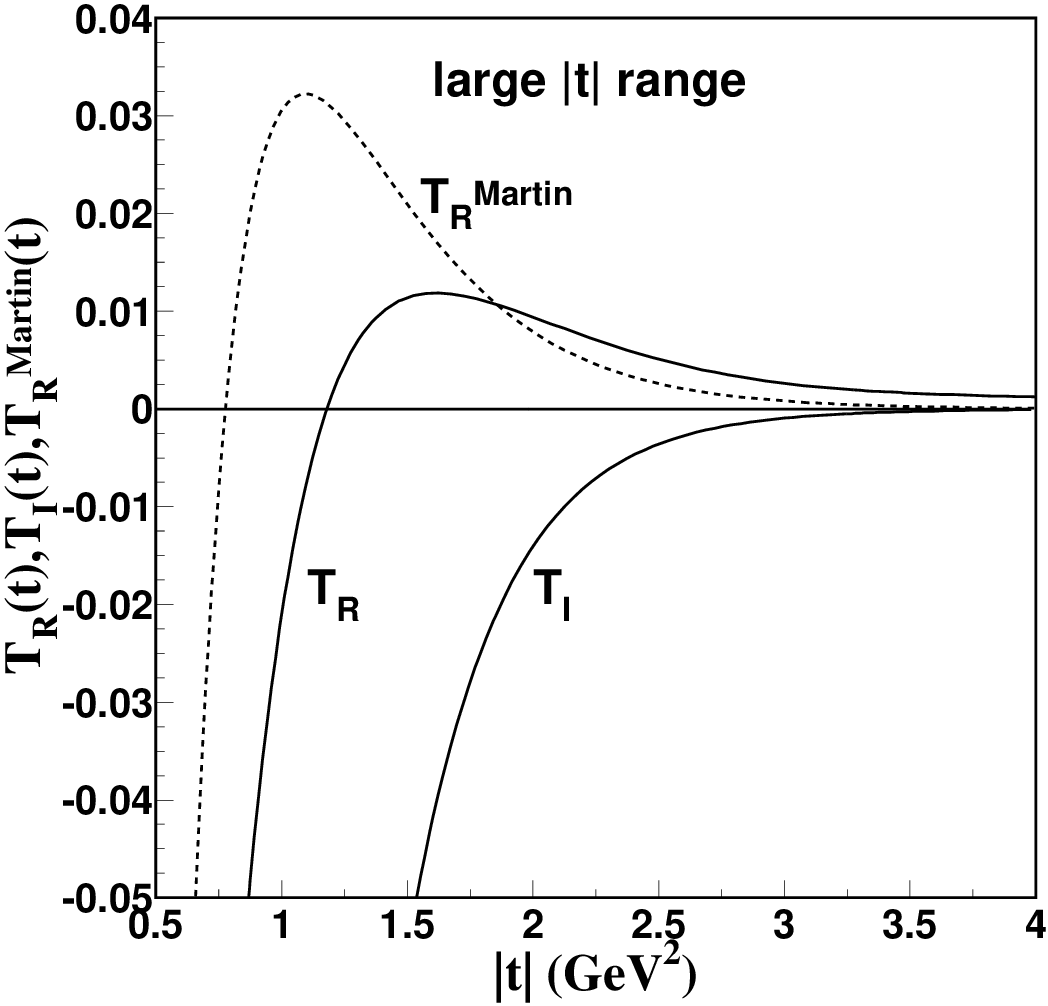} 
        \caption { Martin's Real Part Formula.
      a) $|t|$ dependence of the real part of elastic amplitude calculated with
   Martin's Formula $T^{\rm Martin}_R(t)$ using the imaginary part $T_I(t)$ of KFK model, 
compared with $T_R(t)$, both normalized to 1 at $|t|=0$; ~ 
      b)large $|t|$  behaviour of $T_R(|t|)$  and $T_I(|t|)$ of KFK calculation compared with 
  the prediction $T_R^{\rm Martin}(|t|)$ from Martin's Formula using same $T_I(|t|)$; the
 real amplitudes are positive in both cases, with magnitudes  dominant (slightly in the case of Martin's Formula) over the negative imaginary part.
           \label{Martin_figure}  }  
         \end{figure*}

 \subsection{ BSW and Selyugin's  HEGS  models \label{BSW_Selyugin}} 

\begin{figure*}[b]
          \includegraphics[width=8cm]{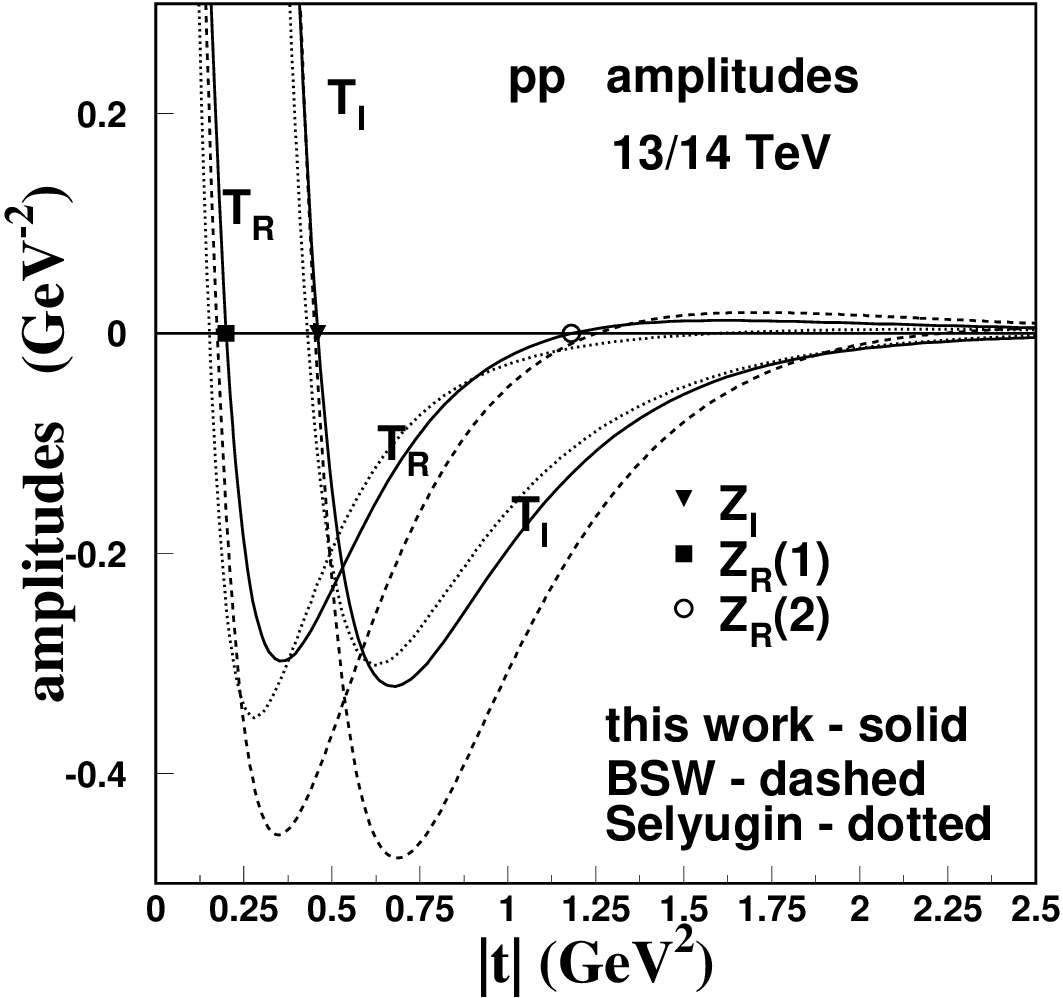} 
\caption {Scattering amplitudes $T_R(t)$ and $T_I(t)$ in comparison 
with other models:  KFK solution of Eqs.(\ref{psi_st},\ref{hadronic_complete}) 
and Table \ref{tableone}
in solid line,  BSW model \cite{BSW} in dashed line and  Selyugin's HEGS model \cite{Selyugin} 
in dotted line. 
The solutions are  similar,  
with zeros in similar  positions, and with dominance of the real part 
(with positive sign) for large $|t|$ in the experimental range. 
  \label{BSW_SEL}  }
\end{figure*}
  
The model proposed by Bourrely, Soffer and Wu (called BSW model) 
 \cite{BSW} gives explicitly the full $s,t$ dependence of the elastic  
 amplitudes and is appropriate for the comparison with the
calculations in KFK.  
 The structure of the pp and ${\rm p\bar p }$ interactions  
studied by O. Selyugin \cite{Selyugin}, based on the analysis  of 
different sets of Parton Distribution Functions and introducing 
t-dependence in the Generalized Parton Distributions, called HEGS model
by the author,  
 gives good representation of $d\sigma/dt$ data for  large 
energy range, predicting  the LHC experiment at 13 TeV.  
 Fig.\ref{BSW_SEL}  shows the   dependences of 
 the   amplitudes predicted by these two models for 13 and  14 TeV 
several years before the experiments.      
The similarity of both BSW and HEGS models  
with present  KFK calculations  in 
the forms of the amplitudes  reinforces  the expectation of the 
present work, that aims at a realistic identification of the terms of 
the complex elastic amplitude. 

\subsection{Models on the space structure of the proton \label{diversos} }   
 
 Recently, Cs\"orgo, Pasechnik and Ster \cite{Csorgo} introduced the statistical
analysis of L\'evy imaging method to extract the information of the colliding proton
structure in a model-independent way and quantify its inelasticity profile in $b$ space, 
obtaining $d\sigma_{\rm inel}/db$ as function of $b$. Comparing the results for 
different energies, they claim that a possible emergence of the "proton hollowness"  
(or equivalently "black ring") 
effect at 13 TeV. Note that their inelastic profile function
$d\sigma_{\rm inel}/db$ is practically identical with our results shown in
Fig.\ref{cross-b-50}.  
 The claimed "hollowness"  is also found in our  $d\sigma_{\rm inel}/db$. although its  
location and intensity are smaller.
In terms of their parameters $H=\exp(-2 \chi_I(0))$ 
and $h=H-\exp({-2 \chi_I(b_{\rm peak})})$,
where $b_{\rm peak}$ is the position where  $\chi_I$
becomes maximum, we have 
$$ b_{\rm  peak} \simeq 0.24 ~{\rm fm} ~,~ H \simeq 0.00346  ~,~h \simeq 0.00033 ~, $$    
compared to the corresponding values in \cite{Csorgo}
 $$ b_{\rm peak} \simeq 0.4 ~ {\rm fm} ~ , ~ H \simeq 0.0085 ~ , ~ h \simeq  0.0058 ~ . $$  
As shown in Fig.\ref{cross-b-50}, 
our analysis predicts that at 50 TeV  this "hollowness" becomes much more enhanced.    

   Similar conjecture of the existence of a layer-structure in the proton,    
revealed  in pp scattering at high energies, based on the observation that 
 there is a range of nearly linear behaviour in $d\sigma/dt$, 
 is  discussed by I.M. Dremin  \cite{Dremin} (and references therein for 
related work). 
In contrast to the above mentioned \cite{Csorgo}   approach, this work deals with the 
elastic profile. The author claims that the enhancement of elastic component 
for  large $|t|$  indicates a hard internal layer in the proton structure. This 
observation also qualitatively agrees with our results, where the elastic profile 
at 13 TeV shows a significant enhancement near $b=0$. As mentioned before, our prediction 
for 50 GeV shows much more clearly the "hard core" structure of the elastic profile 
function for central collisions. Unfortunately, a direct quantitative comparison 
of this work \cite{Dremin} with our result is not available.


 \subsection{Phillips-Barger potential model \label{PB} } 

A paper by V.P. Gon\c calves and P.V.R.G. Silva 
\cite{PV}   uses the formula for the complex amplitude
   based on the Phillips-Barger potential  model  \cite{Pancheri_2,Fagundes}  
\begin{equation}
A_{PB}(t)=i\big[ \frac{1}{(1-t/t_0)^4}\sqrt{A} {\rm e}^{(Bt/2)}+{\rm e}^{i\phi}\sqrt{C}
      {\rm e}^{Dt/2} \big]
\label{PB_eq}
\end{equation}
to parametrize  $d\sigma/dt$  at several energies for the full $|t|$ range.  
 With six free parameters, the 13 TeV data (398 points) are fitted with $\chi^2=6.30$
with statistical errors only. This value looks similar to our value 5.186 for 428 points 
in Table \ref{tabletwo}.  The real part in the amplitude in Eq.(\ref{PB_eq}) 
has  a pure exponential form, without zero, and is  very small in magnitude for all $|t|$, 
   with a value at the origin  $\rho=0.02$. We understand the the treatment of the 
real part in the  framework of this model  is not simple  \cite{Pancheri_2}.

In most  models the  range of transition from  $|t| \approx 2.5 \GeV^2$ 
to the perturbative tail stays   somewhat outside their treatments, indicating need 
of special investigation of this region, and also of more  precise  measurements.

\section{  Final  Comments \label{CONCLUSIONS} }

Elastic scattering is described by one single complex function depending on
two kinetic variables and it is natural to expect that investigations may lead
to explicit and hopefully realistic (compatible with data and with any model independent   
information)  expressions for both parts of this function, as 
is attempted in the present work. 
Besides the $|t|$  amplitudes extracted from data in direct analytical form,  
  the   impact parameter representation  is also explicitly given together with their
eikonal representation,  
so that unitarity  can be studied and controlled, in addition to providing
 physically intuitive images. 
 We believe that the regularity in the energy dependence previously studied \cite{KFK_3} 
and reviewed in Sec. \ref{energy}  adds  reliability to our proposal.

Characteristic features of the disentanglement of the amplitudes here proposed 
are the two zeros of the real part, and
the single zero of the imaginary part.  
Interesting support in this respect comes from 
 the qualitative agreement of the real part in KFK with the prediction from
Martin's Real Part Formula shown in Fig.\ref{Martin_figure},  
with the zeros and the dominant positive real part for large $|t|$.
Since very precise representation of the data is obtained in this work, the results  
suggest bridges between experiments and amplitudes that may serve as 
reference for  other models.

The interplay of the imaginary and real amplitudes
at mid values of $|t|$ is responsible for the dip-bump structure of the
differential cross section.  
 For large $|t|$ the perturbative term of the real 
part is dominant, while at small $|t|$ the imaginary non-perturbative term 
is stronger, occupying about 75 \% of the cross section.

The   Yukawa-like behaviour  for large $b$ of the profile function derived from 
the  loop-loop interaction in the Stochastic Vacuum Model, that is incorporated in  
the input amplitudes of KFK, is present  in   
treatments of the pp interaction through Wilson loop  correlation functions.

In the present analysis, we also studied the possible energy dependence of 
the  model parameters and updated the earlier version \cite{KFK_3}. One new finding is 
that  at $b=0$,  the elastic scattering profile, 
$d\sigma/db_{\rm elas}$ increases with the incident energy very quickly beyond 13 TeV, 
whereas the inelastic profile decreases. These properties are also reported in \cite{Dremin} 
and
\cite{Csorgo}, respectively. The dominance of elastic process at $b=0$ with quick energy 
variation  predicted here, 
 together with the increasing suppression ("hollowness") of inelastic channel, 
  certainly introduces a new clue for the role of proton structure in very high energy collisions. 
Intuitively speaking, at very high energies, the central collision of proton-proton behaves
 as under  a hard-core elastic potential scattering, 
with hard-core repulsion due to  Pauli's exclusion principle. If so, 
naturally we expect that such behavior will not appear  similarly in $\rm { p \bar p }$ scattering. 

Finally we remark that in KFK model, the  
parameters of the real and imaginary parts of the elastic amplitude 
are treated independently. We refer exclusively to the $-t>0 $  half-plane, so that  
we cannot guarantee that final amplitude is analytic when $s$ and $t$ are extended to the complex 
domain. This concern would  impose further constraints, particularly in  extrapolation to higher 
energies.

 Questions of analyticity and crossing symmetry, with explicit inclusion of 
energy dependence, as in frameworks exploring scaling properties \cite{Kohara}, 
require further study.

\section{Acknowledgements}
The authors  wish to thank the Brazilian agencies CNPq, CAPES  and FAPERJ  for financial support.   
Part of the present work was developed under the project INCT-FNA Proc. No. 464898/2014-5.


\end{document}